\documentclass[journal]{IEEEtran}
\usepackage{float}
\usepackage{graphicx}
\usepackage{textcomp}
\usepackage{colortbl}
\usepackage{adjustbox}
\usepackage{caption}
\usepackage{paralist}
\usepackage{tabularx}
\usepackage{import}
\usepackage{algpseudocode,algorithm}
\usepackage{listings}
\usepackage{multirow}
\usepackage{caption}
\usepackage{subcaption}
\usepackage{hyperref}
\usepackage{circuitikz}
\usepackage{pgf}
\usepackage{pgfplots}
\usepackage{xcolor}
\usepackage{bm}
\usepackage{pgfplotstable}
\usepackage[many]{tcolorbox}
\pgfplotsset{compat=1.8}
\usepackage{adjustbox}
\usepackage{changepage}
\usepackage{enumitem}
\usepackage{dblfloatfix}

\usepackage[utf8]{inputenc}

\tcbset{
  highlightbox/.style={
    colback=gray!10,              
    colframe=gray!50,             
    boxrule=0.4pt,                
    arc=5pt,                      
    left=5pt,right=5pt,top=3pt,bottom=3pt, 
    enhanced,                     
  }
}

\newlength{\bubblesep}
\newlength{\bubblewidth}
\AtBeginDocument{\setlength{\bubblewidth}{.75\textwidth}}
\definecolor{bubblegreen}{RGB}{103,184,104}
\definecolor{bubblegray}{RGB}{241,240,240}

\newcommand{\bubble}[4]{%
  \tcbox[
    on line,
    arc=4.5mm,
    colback=#1,
    colframe=#1,
    tcbox width=auto limited, 
    width=\columnwidth, 
    #2,
  ]{\color{#3}#4}%
}

\ExplSyntaxOn
\seq_new:N \l__ooker_bubbles_seq
\tl_new:N \l__ooker_bubbles_first_tl
\tl_new:N \l__ooker_bubbles_last_tl

\NewEnviron{rightbubbles}
 {
  \begin{flushright}
  \sffamily
  \seq_set_split:NnV \l__ooker_bubbles_seq { \par } \BODY
  \int_compare:nTF { \seq_count:N \l__ooker_bubbles_seq < 2 }
   {
    \bubble{bubblegreen}{rounded~corners}{white}{\BODY}\par
   }
   {
    \seq_pop_left:NN \l__ooker_bubbles_seq \l__ooker_bubbles_first_tl
    \seq_pop_right:NN \l__ooker_bubbles_seq \l__ooker_bubbles_last_tl
    \bubble{bubblegreen}{sharp~corners=southeast}{white}{\l__ooker_bubbles_first_tl}
    \par\nointerlineskip
    \addvspace{\bubblesep}
    \seq_map_inline:Nn \l__ooker_bubbles_seq
     {
      \bubble{bubblegreen}{sharp~corners=east}{white}{##1}
      \par\nointerlineskip
      \addvspace{\bubblesep}
     }
    \bubble{bubblegreen}{sharp~corners=northeast}{white}{\l__ooker_bubbles_last_tl}
    \par
   }
   \end{flushright}
 }
\NewEnviron{leftbubbles}
 {
  \begin{flushleft}
  \sffamily
  \seq_set_split:NnV \l__ooker_bubbles_seq { \par } \BODY
  \int_compare:nTF { \seq_count:N \l__ooker_bubbles_seq < 2 }
   {
    \bubble{bubblegray}{rounded~corners}{black}{\BODY}\par
   }
   {
    \seq_pop_left:NN \l__ooker_bubbles_seq \l__ooker_bubbles_first_tl
    \seq_pop_right:NN \l__ooker_bubbles_seq \l__ooker_bubbles_last_tl
    \bubble{bubblegray}{sharp~corners=southwest}{black}{\l__ooker_bubbles_first_tl}
    \par\nointerlineskip
    \addvspace{\bubblesep}
    \seq_map_inline:Nn \l__ooker_bubbles_seq
     {
      \bubble{bubblegray}{sharp~corners=west}{black}{##1}
      \par\nointerlineskip
      \addvspace{\bubblesep}
     }
    \bubble{bubblegray}{sharp~corners=northwest}{black}{\l__ooker_bubbles_last_tl}\par
   }
  \end{flushleft}
 }
\ExplSyntaxOff

\ifCLASSOPTIONcompsoc
  \usepackage[nocompress]{cite}
\else
  \usepackage{cite}
\fi

\ifCLASSINFOpdf
\else
\fi
\begin{document}
%
\title{SleepWalk: Exploiting Context Switching and Residual Power for Physical Side-Channel Attacks}

\author{Sahan~Sanjaya,~\IEEEmembership{Student Member,~IEEE,} Aruna~Jayasena,~\IEEEmembership{Student Member,~IEEE,} and~Prabhat~Mishra,~\IEEEmembership{Fellow,~IEEE,}

\IEEEcompsocitemizethanks{\IEEEcompsocthanksitem S. Sanjaya, A. Jayasena and P. Mishra are with the Department of Computer \& Information Science \& Engineering, University of Florida, USA. Email: \{ssanjaya, arunajayasena, prabhat\}@ufl.edu}
}

\markboth{Journal of \LaTeX\ Class Files,~Vol.~14, No.~8, August~2021}%
{Shell \MakeLowercase{\textit{et al.}}: A Sample Article Using IEEEtran.cls for IEEE Journals}


\maketitle

\begin{abstract}
Context switching is utilized by operating systems to change the execution context between application programs. It involves saving and restoring the states of multiple registers and performing a pipeline flush to remove any pre-fetched instructions, leading to a higher instantaneous power consumption compared to typical program execution. In this paper, we introduce a physical power side-channel leakage source that exploits the power spike observed during a context switch, triggered by the inbuilt \texttt{sleep} function of the system kernel. We observed that this power spike directly correlates with both the power consumption during context switching and the residual power consumption of the previously executed program. Notably, the persistence of residual power signatures from previous workloads extends the scope of this side-channel beyond extracting the data in registers during the context switch. Unlike traditional approaches that require analyzing full power traces, applying complex preprocessing, or relying on external synchronization triggers, this novel technique leverages only the amplitude of a single power spike, significantly simplifying the attack. We developed a power model to illustrate the feasibility of mounting end-to-end side-channel attacks using the \texttt{sleep}-induced power spikes. 
Experimental evaluation demonstrates that our framework can successfully perform cryptographic key recovery for both AES and SIKE implementations on Broadcom BCM2711.
\end{abstract}


%
\IEEEpeerreviewmaketitle

\pagestyle{plain}

\section{Introduction}\label{sec:introduction}


Context switching is utilized by operating systems to enable switching between tasks (application programs)~\cite{li2007quantifying, so2lecture3}. The basic idea behind context switching is to save (store) the state of a task, such as the values stored in registers and other resources, before switching to another task. When the original task is resumed, the previously saved values are restored (load), allowing the task to continue from where it left off. 
However, the concentrated load and store operations associated with context switching lead to higher power consumption due to frequent state transitions between `0' and `1' inside the chip, memory, and buses~\cite{jayasena2024evilcs}. Consequently, this behavior creates a significant attack surface that can be exploited to extract secret information through physical power side-channel attacks~\cite{kocher1999differential}.

\begin{figure}[htp]
    \centering
    \begin{subfigure}{0.49\columnwidth}
        \centering
        \vspace{-0.04in}
        \include{images/intro_traditional}
        \vspace{-0.40in}
        \caption{Traditional power trace.}
        \label{fig:traditional}
    \end{subfigure}
    \hfill
    \begin{subfigure}{0.49\columnwidth}
        \centering
         \vspace{-0.1in}
         \include{images/intro-sleepwalk}
        \vspace{-0.33in}
        \caption{\textit{SleepWalk} power trace.}
        \label{fig:proposed}
    \end{subfigure}

    \begin{subfigure}{\columnwidth}
        \centering
        \scriptsize
        \includegraphics[width=\textwidth]{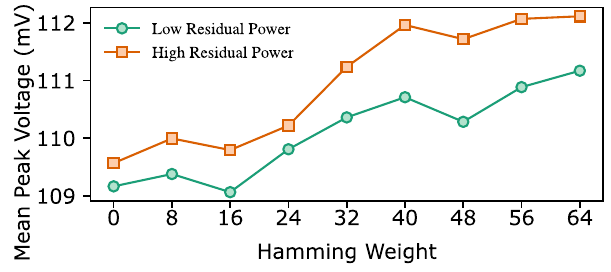}
        \vspace{-0.1in}
        \caption{Effect of the \textbf{context-switching power signature} (green line) and the \textbf{residual power signature} (represented by the power difference between the green and orange lines).}
        \label{fig:residual_and_context}
    \end{subfigure}
    
    \caption{We demonstrate the existence of a \texttt{sleep}-induced power spike (Figure~\ref{fig:proposed}), which directly correlates with both the residual power consumption of the previously executed program and the power consumption during the context switch (Figure~\ref{fig:residual_and_context}). Unlike traditional power analysis (Figure~\ref{fig:traditional}), this side-channel eliminates the need to analyze the entire power trace, perform advanced preprocessing, or use external triggers for trace synchronization.}
    \vspace{-0.2in}
    \label{fig:func_with_without}
\end{figure}

\subsection{Sleep-Induced Power Side-Channel} \label{subsec:intro-sleepwalk}
\begin{figure}[htp]
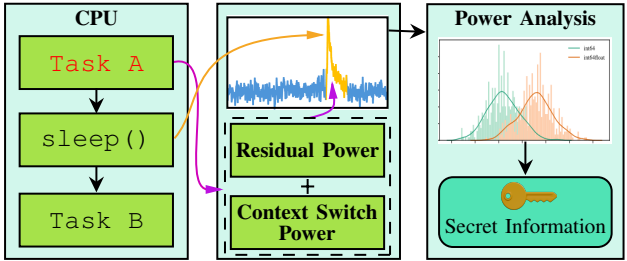

    \centering
    \include{images/Sleepwalk_overview_updated}
    \vspace{-0.3in}
    \caption{Overview of \textit{SleepWalk} power side-channel analysis, which extracts residual and context-switch power signatures from a single \texttt{sleep}-induced power spike to recover secret information from the victim application (\texttt{Task A}).}
    \label{fig:sleepwalk_overview}
    \vspace{-0.2in}
\end{figure}

In this paper, we present a novel power side-channel vulnerability, referred as \textit{SleepWalk}, that leverages both context switching and residual power signatures to extract sensitive information. Although controlled context switches can be induced using various techniques~\cite{zhu2025controlled}, our approach primarily utilizes the operating system kernel’s built-in \texttt{sleep} function, which inherently triggers a context switch. A distinctive power spike is observed at the onset of this transition, as illustrated in Figure~\ref{fig:proposed}.
Unlike traditional power side-channel analysis (as shown in Figure~\ref{fig:traditional}), our approach relies solely on the amplitude of a single sampling point. This significantly reduces complexity by eliminating the need for high-end equipment, advanced preprocessing techniques such as trace alignment, or the extraction of high-dimensional features. Furthermore, the distinct and consistent nature of the observed power spike removes the need for external triggers to synchronize computation with trace collection.
As illustrated in Figure~\ref{fig:residual_and_context}, this power spike is influenced by two key factors. The first is the \textbf{context-switching power signature} (depicted by the green line), which reflects the impact of the Hamming weight of register data at the moment the context switch occurs. The second is the \textbf{residual power signature}, representing the power footprint of instructions executed before the context switch. The difference between the green and orange lines in Figure~\ref{fig:residual_and_context} highlights the contribution of residual power, where the orange line, associated with high residual power, shows a greater amplitude than the green line, which corresponds to low residual power. 
Based on these observations, we introduce the \textit{SleepWalk} side-channel vulnerability that exploits the \texttt{sleep}-induced power spike. Specifically, we demonstrate its significant potential as a security threat by extracting secret information from victim applications, as illustrated in Figure~\ref{fig:sleepwalk_overview}.


\color{black}

\subsection{Contributions}
\label{sec:contributions}


To demonstrate the potential of the \textit{SleepWalk} side-channel, we explore a power model that enables practical side-channel attacks. Using this model, we showcase two attacks: key extraction from post-quantum cryptographic algorithm SIKE, and key recovery from symmetric AES implementations. This paper makes the following key contributions:
\begin{compactitem}


   \item \textbf{\textit{SleepWalk} Side-Channel Vulnerability}: We introduce \textit{SleepWalk}, a novel power side-channel vulnerability that exploits a distinctive power spike generated during context switches invoked by the \texttt{sleep} function. This spike captures both context-switch and residual power signatures, revealing data-dependent behavior.

\item \textbf{Residual Power Signature Analysis}: We develop a power model that characterizes the \texttt{sleep}-induced spike by isolating the contributions of context-switch power and residual power. We analyze how factors such as instruction count, data values, and register states affect the residual power signature.

\item \textbf{\textit{SleepWalk}-based Side-Channel Attacks}: Leveraging our power model, we demonstrate practical side-channel attacks, including cryptographic key recovery. \textit{SleepWalk} enables a single-point leakage model, eliminating the need for trace alignment, external triggers, or complex pre-processing required in power analysis.
    
    
    \item \textbf{Key Recovery Attack on SIKE}: We perform a key recovery attack on the full secret key from the CIRCL SIKE implementation~\cite{armando2019introducing} using \textit{SleepWalk}. 
    We employ a probability density distribution-based classification to successfully extract the full cryptographic key, showcasing the efficacy of this side-channel attack even against hardened implementations. 
    
    \item \textbf{Key Recovery Attack on AES}: We perform a chosen-plaintext attack on AES-128~\cite{dworkin2001advanced}, specifically targeting the final round secret key. We demonstrate that \textit{SleepWalk} can be leveraged to identify processed data at byte-level granularity. Using this approach, we successfully recovered 10 out of 16 bytes of the final round key. Although this result does not outperform existing power side-channel attacks in terms of success rate, it shows that \textit{SleepWalk} can reduce the key recovery complexity from $2^{128}$ to $2^{48}$ using only a single power trace point.

\end{compactitem}

\vspace{0.05in}
To the best of our knowledge, this work is the first to introduce the \texttt{sleep}-induced power spike as a side-channel source for performing key extraction attacks on off-the-shelf hardware.
The remainder of this paper is organized as follows. Section~\ref{sec:background} provides relevant background and surveys related efforts. Section~\ref{sec:leakage_method} introduces a new power side-channel. In Section~\ref{sec:leakage_model}, we develop a power model to show that there is a correlation between the \textit{SleepWalk} side-channel and the internal states of a device during the execution of application programs. 
Section~\ref{sec:sike} and~\ref{sec:aes} demonstrate key recovery attacks on SIKE and AES implementations, respectively.  Section~\ref{sec:conclusion} concludes the paper.

\section{Background and Related work}\label{sec:background}

We first introduce power side-channel analysis and context switching. Next, we survey related efforts. 
Finally, we outline our contributions in terms of the novelty of the side-channel source, analysis technique, and the findings.

\vspace{-0.05in}
\subsection{Power Side-Channel Analysis}

\begin{figure}[htp]
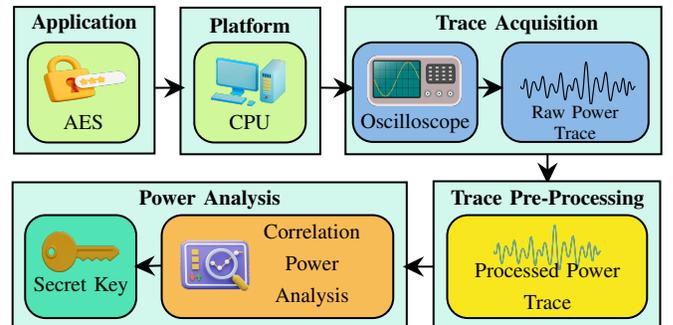

    \centering
    \include{images/PSC_attack_flow}
    \vspace{-0.4in}
    \caption{Overview of Power Side-Channel analysis}
    \label{fig:powerside-channel}
\end{figure}

In recent years, numerous vulnerabilities have been discovered that exploit the power signature of the device revealing internal secrets. Figure~\ref{fig:powerside-channel} illustrates the fundamental process of performing a power side-channel attack. This process typically consists of three main phases: trace collection, pre-processing, and analysis. 

\noindent\textbf{Trace Collection}:
The trace collection phase involves capturing power variations corresponding to computations occurring within the targeted hardware. The trace can be collected by probing the power lines and measuring the voltage fluctuations across components of the device~\cite{gamaarachchi2018power}.

\vspace{0.05in}
\noindent\textbf{Trace Pre-processing}:
A key challenge in this step is ensuring synchronization between the acquired traces and the targeted computation. To achieve this, adversaries often use an external trigger signal to isolate the relevant portion of the trace that corresponds to the computation of interest~\cite{guillaume2022virtual}. The adversaries must employ equipment capable of capturing a sufficient number of sampling points per trace to maximize the chances of a successful attack. The analysis phase may utilize various trace characteristics, such as amplitude or frequency response, to extract sensitive information. To enhance signal quality, techniques such as averaging and filtering improve the signal-to-noise ratio (SNR)~\cite{moos2019static, gu2023trace}, while alignment methods optimize trace consistency.

\vspace{0.05in}
\noindent\textbf{Power Analysis}:
The pre-processed traces are used to perform either a profiled or non-profiled attack. For non-profiled attacks,  such as Simple Power Analysis (SPA)~\cite{kocher1996timing,ahmadi2023shield}, Differential Power Analysis (DPA)~\cite{kocher1999differential,aysu2018binary}, or Correlation Power Analysis (CPA)~\cite{brier2004correlation,benhadjyoussef2021power}, time-domain power traces with sufficient sampling points are adequate. In contrast, profiled attacks rely on a profiling model, such as a machine learning (ML) model~\cite{gao2024deeptheft,ahmed2023deep, yang2016inferring}, and require additional features, such as minimum, maximum, standard deviation, spectral centroid, spectral entropy, spectral irregularity, and spectral spread \cite{matovu2020defensive, chen2017powerful}, to extract the targeted information.

\subsection{Context Switching in Multitasking Systems}\label{subsec:threads}

\begin{figure}[htp]
    \centering
    \footnotesize
    \vspace{-0.2in}
    \input{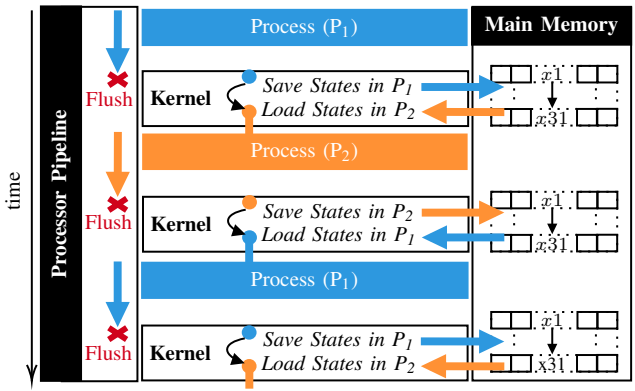}
    \vspace{-0.3in}
    \caption{Context switch and related pipeline flush routines}
    \label{fig:contextsw}
    \vspace{-0.1in}
\end{figure}

\vspace{0.1in}
\noindent\textbf{Context Switch}:
During a context switch, the operating system kernel must execute a series of operations to ensure a smooth transition between processes, as illustrated in Figure~\ref{fig:contextsw}. First, it performs a sequence of memory writes (\texttt{store}) to save the current process's register state into memory, preserving its execution context. Then, it executes a sequence of memory reads (\texttt{load}) to restore the register state of the next process, allowing it to resume execution from where it was last interrupted.

\vspace{0.1in}
\noindent\textbf{Pipeline Flush}:
During a context switch, the processor must flush the pipeline to clear any pre-fetched instructions from the previous process. Moreover, modern processors with branch prediction and speculative execution must reset these mechanisms to avoid executing incorrect or unauthorized instructions that were prefetched from the previous process. The Translation Lookaside Buffer (TLB), which holds virtual-to-physical address mappings, may require flushing unless address space identifier-based TLB entries allow selective retention. If the processor maintains an instruction prefetch buffer, it should be cleared to ensure that the processor only executes the instructions from the new process.

\subsection{Related Work}
\label{subsec:related}

We survey existing efforts that utilize power side-channel analysis to reconstruct the information that are carried out within multi-processing environments. 

\vspace{0.05in}
\noindent\textbf{Software-based Power Trace Collection}:
Intel Running Average Power Limit (RAPL)  has been utilized by many existing efforts to demonstrate power side-channel attacks~\cite{gao2024deeptheft, lipp2021platypus, zhang2021red}. DeepTheft~\cite{gao2024deeptheft} is an end-to-end attack that accurately reconstructs complex Deep Neural Network (DNN) architectures by exploiting the RAPL interface on x86 processors. 
PLATYPUS~\cite{lipp2021platypus} is another software-based power side-channel attack that can successfully extract cryptographic keys. 
Zhang et al. demonstrated that unprivileged access to RAPL allows attackers to infer power consumption patterns, leading to leakage of sensitive information~\cite{zhang2021red}. Hertzbleed~\cite{wang2022hertzbleed} attack transforms power side-channel vulnerabilities into remote timing attacks, allowing attackers to infer cryptographic keys. 

\vspace{0.05in}
\noindent\textbf{Hardware-based External Power Monitoring}:
There are promising research efforts that utilize physical measurements to capture information about the power trace. \cite{moos2017static} conducted a static power side-channel analysis on an ASIC implementation of a threshold implementation prototype chip. Their study demonstrated that even hardware designs with provable first-order security are susceptible to information leakage through static power consumption, highlighting the necessity for enhanced countermeasures in cryptographic hardware. \cite{sanjaya2025information} explored the Physical Layer Supply Voltage Coupling (PSVC) vulnerability. This study revealed that the power side-channel signature of a physical processor core can propagate across design boundaries via the power rails. 
There are various efforts for power side-channel evaluations on pre-silicon hardware designs~\cite{jayasena2024evilcs, zhang2021psc, pundir2022power, jayasena2025ciseleaks, he2019rtl,jayasena2023tvla}. For example, EvilCS~\cite{jayasena2024evilcs} introduced a framework for assessing information leakage during context switches within security enclave kernels. 

\vspace{0.05in}
\noindent\textbf{Limitations of Existing Efforts: }
Existing power side-channel attacks are conducted either by monitoring software interfaces or through direct physical measurements. These attacks typically rely on capturing the entire power trace \cite{kocher1999differential,aysu2018binary, brier2004correlation,benhadjyoussef2021power, sanjaya2025information} or several sampling points \cite{lipp2021platypus, kogler2023collide, lipp2022amd} for analysis. To enhance trace resolution, they employ advanced processing techniques and utilize external triggers for precise trace alignment. These requirements are expensive in terms of both time and cost, and therefore, such attacks may not be feasible in many application scenarios.
\section{\textit{SleepWalk} Power Leakage Channel}
\label{sec:leakage_method}
\
In this section, we show the existence of the \texttt{sleep}-induced power spike that serves as the basis for the \textit{SleepWalk} side-channel. 
First, we describe the experimental setup. Next, we analyze the observations made during the power spike induced by the internal \texttt{sleep} function. Finally, we perform an experiment to investigate how different workloads affect the behavior of the observed power spike. 

    


\subsection{Experimental Setup}\label{subsec:setup}


The experiments are conducted on a  Raspberry Pi 4 model B single board computer (SBC) with a BCM2711 System-on-Chip (SoC) that integrated an ARM Cortex-A72 CPU. 
We used the GNU/Linux 12 Debian (Bookworm) operating system with kernel version 6.6 with the GLIBC version of 2.36. For trace acquisition, we used a Keysight DSOX1102G oscilloscope~\cite{Keysight}. The oscilloscope probe was connected to the 5V and GND pins of the Raspberry Pi device. The oscilloscope was controlled using the Virtual Instrument Software Architecture (VISA) protocol to capture the traces. A simple running average with a window size of 10 was applied to smooth the traces before finding the peak values. We have used Python scripts to automate the trace collection process and to facilitate peak value extraction. The captured peak values are then stored for further analysis according to the requirements of the specific attack, including cryptographic key recovery attacks on SIKE (Section~\ref{sec:sike}) and AES (Section~\ref{sec:aes}) algorithms.



\subsection{Sleep and Power Spike}

In order to illustrate the process of creating the \texttt{sleep}-induced power side-channel, we use two programs \texttt{Task A} and \texttt{Task B} that execute different sequences of assembly instructions. We first describe the observations we made during the process of introducing a \texttt{sleep} function call from the operating system while these functions are executed. Next, we discuss the reason behind these observations with respect to the analysis of the kernel and hardware routines.

\vspace{0.1in}
\noindent\textbf{Observation}:
Figure~\ref{fig:func_with_without} illustrates the power trace of the program executing two tasks: \texttt{Task A} and \texttt{Task B}. First, we collected the power trace of running the two tasks one after the other, as shown in Figure~\ref{fig:traditional}. 
Next, we inserted the \texttt{sleep} function call between these two tasks, as shown in Figure~\ref{fig:proposed}. We observe that adding a \texttt{sleep} function between the two tasks generates a distinct power spike in the captured trace. After reverse-engineering the \texttt{sleep} function’s implementation, we identified that the power spike that we observed is tied to the power signature of a context switch. Next, we explain the reverse-engineered analysis supporting this claim.

    
    

\begin{figure}[htp]
\centering
\vspace{-0.1in}
\footnotesize
\tikzset{every picture/.style={line width=0.75pt}} 

\begin{tikzpicture}[x=0.75pt,y=0.75pt,yscale=-0.65,xscale=0.6]

\draw    (95.67,77.36) -- (95.67,102.3) ;
\draw [shift={(95.67,105.3)}, rotate = 270] [fill={rgb, 255:red, 0; green, 0; blue, 0 }  ][line width=0.08]  [draw opacity=0] (8.93,-4.29) -- (0,0) -- (8.93,4.29) -- cycle    ;
\draw    (97,111.36) -- (179,111.36) ;
\draw [shift={(182,111.36)}, rotate = 180] [fill={rgb, 255:red, 0; green, 0; blue, 0 }  ][line width=0.08]  [draw opacity=0] (8.93,-4.29) -- (0,0) -- (8.93,4.29) -- cycle    ;
\draw    (251,136.8) -- (251,162.3) ;
\draw [shift={(251,165.3)}, rotate = 270] [fill={rgb, 255:red, 0; green, 0; blue, 0 }  ][line width=0.08]  [draw opacity=0] (8.93,-4.29) -- (0,0) -- (8.93,4.29) -- cycle    ;
\draw   (179,167.36) -- (322.6,167.36) -- (322.6,199.36) -- (179,199.36) -- cycle ;

\draw    (323,183.36) -- (345,183.36) ;
\draw [shift={(348,183.36)}, rotate = 180] [fill={rgb, 255:red, 0; green, 0; blue, 0 }  ][line width=0.08]  [draw opacity=0] (8.93,-4.29) -- (0,0) -- (8.93,4.29) -- cycle    ;
\draw    (419,207.3) -- (419,226.3) ;
\draw    (419,226.3) -- (545.17,226.3) ;
\draw [shift={(548.17,226.3)}, rotate = 180] [fill={rgb, 255:red, 0; green, 0; blue, 0 }  ][line width=0.08]  [draw opacity=0] (8.93,-4.29) -- (0,0) -- (8.93,4.29) -- cycle    ;
\draw    (548.17,236.3) -- (548.17,258.3) ;
\draw [shift={(548.17,261.3)}, rotate = 270] [fill={rgb, 255:red, 0; green, 0; blue, 0 }  ][line width=0.08]  [draw opacity=0] (8.93,-4.29) -- (0,0) -- (8.93,4.29) -- cycle    ;

\draw (250,52.08) node   [align=left] {\begin{minipage}[lt]{46.69pt}\setlength\topsep{0pt}
\begin{center}
{\footnotesize Kernel}
\end{center}

\end{minipage}};
\draw (250,69.36) node   [align=left] {\begin{minipage}[lt]{46.49pt}\setlength\topsep{0pt}
\begin{center}
{\footnotesize \texttt{Task A}}
\end{center}

\end{minipage}};
\draw (419,52.08) node   [align=left] {\begin{minipage}[lt]{46.69pt}\setlength\topsep{0pt}
\begin{center}
{\footnotesize Kernel}
\end{center}

\end{minipage}};
\draw (419,69.36) node   [align=left] {\begin{minipage}[lt]{46.49pt}\setlength\topsep{0pt}
\begin{center}
{\footnotesize \texttt{Task B}}
\end{center}

\end{minipage}};
\draw (530,52.08) node   [align=left] {\begin{minipage}[lt]{46.69pt}\setlength\topsep{0pt}
\begin{center}
{\footnotesize User space}
\end{center}

\end{minipage}};
\draw (530,69.36) node   [align=left] {\begin{minipage}[lt]{46.49pt}\setlength\topsep{0pt}
\begin{center}
{\footnotesize \texttt{Task B}}
\end{center}

\end{minipage}};
\draw (95.67,52.08) node   [align=left] {\begin{minipage}[lt]{46.69pt}\setlength\topsep{0pt}
\begin{center}
{\footnotesize User space}
\end{center}

\end{minipage}};
\draw (95.67,69.36) node   [align=left] {\begin{minipage}[lt]{46.49pt}\setlength\topsep{0pt}
\begin{center}
{\footnotesize \texttt{Task A}}
\end{center}

\end{minipage}};
\draw (250.8,183.36) node   [align=left] {\begin{minipage}[lt]{97.65pt}\setlength\topsep{0pt}
\begin{center}
{\footnotesize Context switch}
\end{center}

\end{minipage}};
\draw (296,149.86) node   [align=left] {\begin{minipage}[lt]{46.69pt}\setlength\topsep{0pt}
\begin{center}
{\footnotesize schedule()}
\end{center}

\end{minipage}};
\draw (484.67,237.86) node   [align=left] {\begin{minipage}[lt]{46.69pt}\setlength\topsep{0pt}
\begin{center}
{\footnotesize exit syscall}
\end{center}

\end{minipage}};
\draw    (180.5,87.3) -- (321.5,87.3) -- (321.5,136.3) -- (180.5,136.3) -- cycle  ;
\draw (251,111.8) node   [align=left] {\begin{minipage}[lt]{93.57pt}\setlength\topsep{0pt}
\begin{center}
{\footnotesize Save user registers }\\{\footnotesize to kernel stack }
\end{center}

\end{minipage}};
\draw    (348.5,159.51) -- (489.5,159.51) -- (489.5,206.51) -- (348.5,206.51) -- cycle  ;
\draw (419,183.01) node   [align=left] {\begin{minipage}[lt]{93.57pt}\setlength\topsep{0pt}
\begin{center}
{\footnotesize Pop user registers }\\{\footnotesize from kernel stack}
\end{center}

\end{minipage}};
\draw (132.9,130.68) node   [align=left] {\begin{minipage}[lt]{33.32pt}\setlength\topsep{0pt}
\begin{center}
{\footnotesize syscall interrupt}
\end{center}

\end{minipage}};

\end{tikzpicture}
\vspace{-0.3in}
\caption{Timing flow of Linux kernel context switch: saving context for \texttt{Task A} and restoring context for \texttt{Task B}.}
\label{fig:linux_context_switch}
\end{figure}
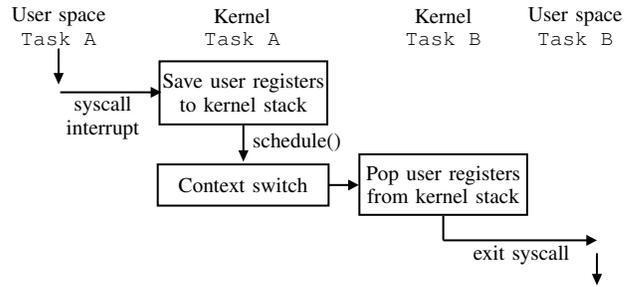

\begin{figure}[ht]
  \centering
  \scriptsize
  \begin{minipage}{0.9\columnwidth}
    \begin{center}
    \begin{lstlisting}[]
sleep
 |->xnanosleep 
    |->nanosleep 
      |->clock_nanosleep 
            |->invoke_syscall
               |->__arm64_sys_clock_nanosleep
                |->common_nsleep
                  |->hrtimer_nanosleep
                    |->do_nanosleep
                      |->schedule
    \end{lstlisting}
    \end{center}
  \end{minipage}
  \vspace{0.2in}
  \caption{\texttt{sleep()} function's function call trace}
  \label{fig:sleep_call_trace}
  \vspace{-0.1in}
\end{figure}

\vspace{0.1in}
\noindent\textbf{Analysis on the Observation}:
Context switching is the process by which the kernel saves the state of the currently running process and restores the state of another process, allowing multiple processes to effectively share a single CPU. This mechanism is important for multitasking, and it is managed by the operating system kernel's scheduler. As shown in Figure~\ref{fig:linux_context_switch}, the \texttt{schedule} function is called when a context switch is required~\cite{so2lecture3}. 
Since \texttt{schedule} is a kernel function, it cannot be invoked directly from user space. However, user-space applications can trigger context switches indirectly through various system calls or operations that cause the kernel to suspend the current process, prompting the scheduler to allocate CPU time to other processes. Figure~\ref{fig:sleep_call_trace} illustrates the function call trace during the execution of the \texttt{sleep} function. It clearly shows that a context switch occurs during the initialization of \texttt{sleep} through the invocation of \texttt{schedule}. Therefore, the power spike observed in Figure~\ref{fig:func_with_without} is expected to reflect the power signature associated with a context switch. To confirm that the spike is indeed caused by the context switch, we conduct an experiment using various context switch invocation methods, including \texttt{sleep}, \texttt{usleep}, \texttt{nanosleep}, \texttt{sched\_yield}, \texttt{pause}, and \texttt{kill} with \texttt{SIGSTOP} signal\cite{linuxmanpage}.

\begin{figure}[htp]
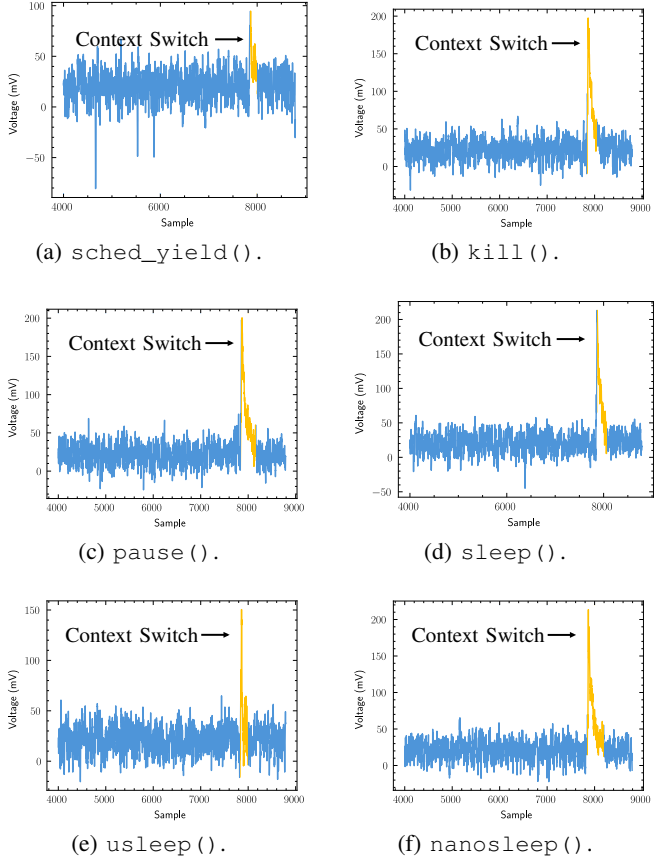

    \centering
    \vspace{-0.1in}
    \begin{subfigure}{0.48\columnwidth}
        \centering
        \vspace{0pt}
        \include{images/Trace5007_colored}
        \vspace{-0.45in}
        \caption{\texttt{sched\_yield()}.}
    \end{subfigure}
    \hfill
    \begin{subfigure}{0.48\columnwidth}
        \centering
        \vspace{0pt}
        \include{images/Trace5008_colored}
        \vspace{-0.45in}
        \caption{\texttt{kill()}.}
    \end{subfigure}
    
    \vspace{0.2cm} 
    
    \begin{subfigure}{0.48\columnwidth}
        \centering
        \vspace{0pt}
        \include{images/Trace5009_colored}
        \vspace{-0.45in}
        \caption{\texttt{pause()}.}
    \end{subfigure}
    \hfill
    \begin{subfigure}{0.48\columnwidth}
        \centering
        \vspace{0pt}
        \include{images/Trace5010_colored}
        \vspace{-0.45in}
        \caption{\texttt{sleep()}.}
    \end{subfigure}

    \vspace{0.2cm} 
    
    \begin{subfigure}{0.48\columnwidth}
        \centering
        \vspace{0pt}
        \include{images/Trace5011_colored}
        \vspace{-0.45in}
        \caption{\texttt{usleep()}.}
    \end{subfigure}
    \hfill
    \begin{subfigure}{0.48\columnwidth}
        \centering
        \vspace{0pt}
        \include{images/Trace5012_colored}
        \vspace{-0.45in}
        \caption{\texttt{nanosleep()}.}
    \end{subfigure}
    
    \caption{Power traces collected during the execution of various user-accessible system calls that trigger a context switch. All traces consistently exhibit a distinctive power spike from the normal execution.}
    \label{fig:diff_context_switch_methods}
    \vspace{-0.15in}
\end{figure}

As illustrated in Figure~\ref{fig:diff_context_switch_methods}, the results confirm that the power spike is caused by the context switch. Among the different methods, \texttt{kill} and \texttt{pause} require a process ID (PID), which limits their applicability in realistic attack scenarios. Additionally, \texttt{sched\_yield} induces a less detectable power spike. In contrast, \texttt{sleep}, \texttt{usleep}, and \texttt{nanosleep} are easier to use and generate more prominent power spikes, making them more suitable for our analysis. As a result, all experiments in this paper utilize context switches induced by the \texttt{sleep} function, hence we named this leakage channel as \textit{SleepWalk}.

\subsection{Distinguishing Programs with \textit{SleepWalk}}

\begin{figure}[htp]
\scriptsize
\centering
\vspace{-0.1in}
\includegraphics[width=0.4\textwidth]{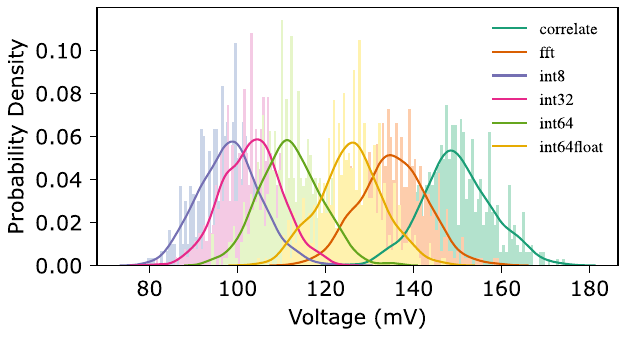}
\vspace{-0.05in}
\caption{Probability density distribution of \texttt{sleep}-induced peak power values for six workloads. The \texttt{sleep} function was inserted at the end of each program, and each program was executed for one second.}
\label{fig:fingerprint_inside_sleep}
\end{figure}

To demonstrate that the \textit{SleepWalk} side channel can be leveraged for side-channel attacks, we performed an initial program-distinguishing experiment. For this experiment, we used the setup described in Section~\ref{subsec:setup} and executed six different application programs from the Stress-NG~\cite{king2017stress} benchmark suite to evaluate how varying workloads affect \texttt{sleep}-induced power spike. Each workload was executed for a fixed duration of one second before invoking the \texttt{sleep} function at the end. The resulting power traces were collected to compute the probability density distribution of the peak power values. Figure~\ref{fig:fingerprint_inside_sleep} illustrates these distributions for each workload. While some overlap exists among the curves, the peak power values still provide distinguishability across workloads. Notably, the \texttt{int8}, \texttt{int32}, and \texttt{int64} workloads use the same computational function but operate on different integer sizes. The ability to differentiate between these three workloads suggests that \textit{SleepWalk} side-channel can also be used to distinguish data or data types used by the previously executed programs.

The key takeaway from this experiment is that there exists a correlation between the \texttt{sleep}-induced power spike and the executed program and its data. Building on this insight, in Section~\ref{sec:leakage_model}, we develop a power model for the \texttt{sleep}-induced power spike to analyze its correlation with the data held in registers and processed by the CPU prior to the context switch. 


\section{\textit{SleepWalk} Leakage Power Model}
\label{sec:leakage_model}

Power side-channel attacks exploit the correlation between the power consumption of a device and the operations performed or data processed internally. To effectively analyze this relationship, power models are employed as theoretical frameworks that approximate the power consumption of a device during computation. These models bridge the gap between observed power traces and the internal states of the device, enabling adversaries to deduce sensitive information such as cryptographic keys.

Existing research employed various power models, including Hamming weight (HWT)~\cite{messerges2000using} and Hamming distance (HD)~\cite{brier2004correlation}. The HWT model estimates the power consumption based on the number of `1's in a binary value. This model is particularly relevant in devices where the power usage correlates strongly with the number of bits set to `1' in a register during computation. In contrast, the HD model measures the power consumption based on the switching between two binary states, i.e., the number of bits that change from one state to another. This model is useful in capturing the dynamic power variations arising from transitions in sequential logic circuits. Beyond the HWT and HD power models, adversaries often develop custom power models to construct an application-specific attack framework~\cite{xiang2020open, kogler2023collide}. In this section, we construct the power model underlying the \textit{SleepWalk} side-channel with controlled experiments on HD and HWT of data.

\subsection{Power Spike Components}
\label{sec:power_model_comp}


As discussed in Section~\ref{subsec:intro-sleepwalk}, the \texttt{sleep}-induced power spike contains two main components: (1) the power consumption associated with the context switch, and (2) the residual power consumption from the previously executed program. While the context switch component directly corresponds to the power usage at the moment of switching and is relatively straightforward to interpret, understanding the residual component requires an in-depth analysis.


\vspace{-0.1in}
\begin{equation}\label{eqn:1}
    P_{\text{total}} = P_{\text{dyn}} + P_{\text{leak}} + P_{\text{sc}}.
    \vspace{-0.1in}
\end{equation}
\begin{equation}\label{eqn:2}
    P_{\text{leak}} \propto V_{dd} \cdot I_{\text{leak}} \quad \text{,} \quad I_{\text{leak}} \propto T^2 e^{-\frac{V_{th}}{nV_T}},
\end{equation}

In CMOS circuits, total power dissipation $P_{\text{total}}$ consists of dynamic power $P_{\text{dyn}}$, leakage power $P_{\text{leak}}$, and short-circuit power $P_{\text{sc}}${\cite{pedram2006thermal}} as illustrated by Equation~\ref{eqn:1}.
Although dynamic power is typically the dominant component in digital CMOS logic, leakage power has become a significant source of power dissipation in modern digital integrated circuits due to increased transistor density~\cite{pedram2006thermal}. This behavior is captured by Equation~\ref{eqn:2}, where $V_{th}$ is the threshold voltage, $n$ is a process-dependent constant, and $V_T = {kT}/{q}$ is the thermal voltage~\cite{liu2007accurate}. According to Equation~\ref{eqn:2}, leakage current ($I_{\text{leak}}$) is proportional to temperature ($T$)  and thus leakage power  ($P_{\text{leak}}$) increases with temperature ($T$)~\cite{liu2007accurate, vassighi2006thermal}.
\begin{equation}\label{eqn:3}
    C_{\text{th}}\,\frac{dT(t)}{dt} = P_{\text{total}} - \frac{T(t) - T_{\text{amb}}}{R_{\text{th}}},
\end{equation}
On-chip temperature change can be modeled using a simple lumped RC thermal network as illustrated in Equation~\ref{eqn:3}, where $C_{\text{th}}$ is the effective thermal capacitance, $R_{\text{th}}$ is the thermal resistance to the ambient, and $T_{\text{amb}}$ is the ambient temperature.
Based on Equations~\ref{eqn:1}, \ref{eqn:2}, and \ref{eqn:3}, we can construct the coupled relationship between power and temperature. Switching activity increases $P_{\text{dyn}}$, which in turn raises total power dissipation and chip temperature. The elevated temperature then further increases $P_{\text{leak}}$, creating a positive feedback loop that amplifies total power. This feedback loop typically converges to a stable power-thermal point~\cite{vassighi2006thermal}.

Based on this power-thermal feedback loop, we identify two key factors that influence steady-state power behavior. The first is switching activity, which directly impacts $P_{\text{dyn}}$ and is modeled using HWT and HD. The second is execution time, which affects temperature buildup; longer execution durations lead to higher temperatures and, consequently, increased leakage power. Execution time is quantified by the number of instructions executed. This interplay between power and thermal effects becomes observable in the \texttt{sleep}-induced power spike as the residual power signature.


To validate the above theoretical analysis about the observed power spike, we conducted four controlled experiments, each designed to isolate a specific factor and examine its relationship to the two key components: context-switch power and residual power. These experiments were guided by the following four hypotheses.

\begin{itemize}
\item Effect of \textbf{HWT of Register Data} corresponds to context-switch power component. (Section~\ref{sec:hwt})
\item Effect of  \textbf{HWT of Processed Data} corresponds to residual power component. (Section~\ref{sec:residual_HW})
\item Effect of \textbf{HD of Processed Data} corresponds to residual power component. (Section~\ref{sec:residual_HD})
\item Effect of  \textbf{Number of Executed Instructions} corresponds to residual power component. (Section~\ref{sec:residual_num_inst})
\end{itemize}

Each data point in these experiments represents the average of 1,000 measured power spikes, collected using the experimental setup detailed in Section~\ref{subsec:setup}.


\begin{figure}[ht]
  \centering
  \scriptsize
    \begin{center} 
    \begin{lstlisting}
const_loop:
    B const_loop
end const_loop:
    // Value with fixed HWT
    X0 = HWT_VALUE
    // Load all GPR with the same value
    MOV X1, X0         
    ::
    MOV X26, X0
    // Make a system call for sleep
    ADR X0, timespec      
    MOV X1, XZR             
    MOV X8, #101              
    SVC #0                    
    \end{lstlisting}
    \end{center}
  \vspace{0.15in}
  \caption{Code for context switch HWT experiment.}
  \label{fig:context_switch_HW}
  \vspace{-0.3in}
\end{figure}

\subsection{Hamming Weight of Data in Registers}
\label{sec:hwt}

We start the experiment with the following hypothesis regarding the data in the registers during the context switch.
\begin{tcolorbox}[highlightbox]
Hypothesis 1: The effect of the Hamming weight of the data in registers has an effect on the \texttt{sleep}-induced power spike. 
\end{tcolorbox}

To analyze the impact of the number of `1's in the data stored in the registers before a context switch, we utilize the program illustrated in Figure~\ref{fig:context_switch_HW}. This program is designed to evaluate the effect of HWT on register values while keeping the other factors constant.
Our execution setup consists of the following two phases:

\begin{itemize}

\item \underline{\textit{Phase 1 [Constant Loop Execution]}}: The first phase contains a loop that provides a reasonable gap between two \texttt{sleep}-induced power spikes. This phase remains unchanged across experiments with varying HWT values. The loop ensures sufficient execution time while maintaining a consistent baseline effect across all HWT experiments.

\item \underline{\textit{Phase 2 [Register Configuration]}}: In the second phase, a value with a fixed HWT (\texttt{HWT\_VALUE}) is loaded into the general-purpose registers to amplify the effect during the context switch. Next, a context switch is triggered using the system call to the kernel's inbuilt \texttt{sleep} function.

\end{itemize}

\begin{figure}[htp]
\centering
\includegraphics[width=0.45\textwidth]{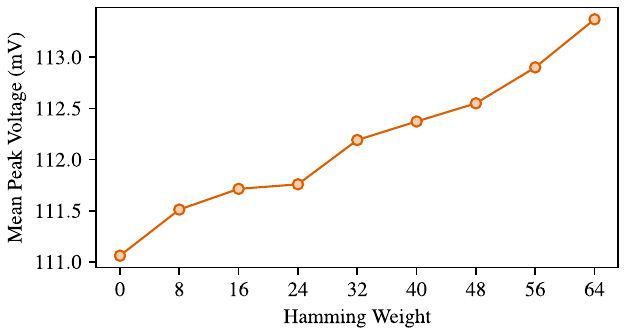}
\vspace{-0.05in}
\caption{Effect of varying the HWT of data in registers (\texttt{HWT\_VALUE}) prior to the context switch. An increase in HWT results in an increase in the mean peak power.}
\label{fig:context_switch_HW_results}
\vspace{-0.15in}
\end{figure}

Through this controlled setup, we are able to analyze the relationship between HWT and the power variations observed during context switching, providing a clear understanding of HWT's impact on this side-channel. 
After collecting the traces, we isolated the peak amplitudes of the traces corresponding to the context switch events. Figure~\ref{fig:context_switch_HW_results} illustrates the results obtained by varying the HWT. The data clearly demonstrates a positive correlation between HWT and mean peak power value, with the peak power increasing as the HWT increases. This behavior confirms our hypothesis 1.

\vspace{0.1in}
\subsection{Hamming Weight of the Processed Data} 
\label{sec:residual_HW}

 We designed this experiment to analyze the effect of the HWT of the processed data prior to a context switch. In order to evaluate this, we create the following hypothesis,
\begin{tcolorbox}[highlightbox]
Hypothesis 2: The effect of the Hamming weight of the data that was processed earlier has an effect on the residual power component of the \texttt{sleep}-induced power spike.
\end{tcolorbox}

\begin{figure}[ht]
  \centering
  \begin{minipage}{0.9\columnwidth}
    \begin{center} 
    \begin{lstlisting}
// Value with fixed HWT
X0 = HWT_VALUE_LOOP 
// Number of loop iterations
X10 = LOOP_ITR
// Load all GPR with the same value
MOV X1, X0         
::
MOV X26, X0
const_loop:
    // Perform OR operations
    OR X2, X1, X0    
    ::
    OR X26, X1, X0
    B const_loop
end const_loop: 
    // Value with fixed HWT 
    X0 = HWT_VALUE 
    // Load all GPR with the same value
    MOV X1, X0         
    ::
    MOV X26, X0 
    // Make a system call for sleep
    ADR X0, timespec      
    MOV X1, XZR             
    MOV X8, #101              
    SVC #0                        
    \end{lstlisting}
    \end{center}
  \end{minipage}
  \vspace{0.2in}
  \caption{Code for residual power HWT experiment}
  \label{fig:residual_HW}
\end{figure}

To observe the effect of the HWT of the data, it is essential to isolate it from the overall power signature. To achieve this, a fixed HWT value (\texttt{HWT\_VALUE\_LOOP}) is first loaded into the general-purpose registers. Next, we introduce an instruction loop that performs \texttt{OR} operations on the values stored in these registers, as illustrated in Figure~\ref{fig:residual_HW}. Since all registers contain the same value, there are no bit transitions during execution, effectively eliminating any influence from the HD of the data. The \texttt{OR} operation is specifically selected because it preserves the HWT while eliminating the effect of HD.
After executing this loop, another fixed HWT value (\texttt{HWT\_VALUE}) is loaded into the general-purpose registers before invoking the \texttt{sleep} function. This setup creates two distinct HWT-related effects: one from the earlier \texttt{OR} operation loop, and another from the register values present at the time of the context switch. As discussed in Section~\ref{sec:hwt}, the HWT of data in the registers during a context switch affects the resulting power spike. Therefore, if that were the only contributing factor, varying \texttt{HWT\_VALUE\_LOOP} while keeping \texttt{HWT\_VALUE} constant should not change the mean power value.

\begin{figure}[htp]
    \centering
    \tiny
    \begin{subfigure}{0.48\columnwidth}
        \centering
        \includegraphics[width=\textwidth]{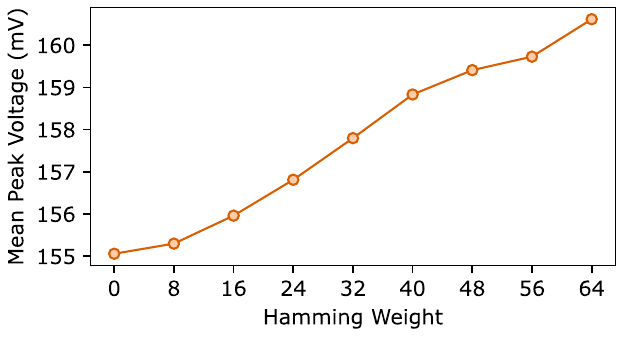}
        \caption{\texttt{HWT\_VALUE} = 0}
    \end{subfigure}
    \hfill
    \begin{subfigure}{0.48\columnwidth}
        \centering
        \includegraphics[width=\textwidth]{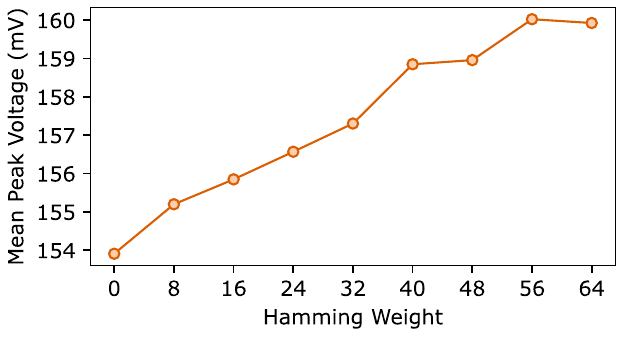}
        \caption{\texttt{HWT\_VALUE} = 16}
    \end{subfigure}
    
    \vspace{0.2cm} 
    
    \begin{subfigure}{0.48\columnwidth}
        \centering
        \includegraphics[width=\textwidth]{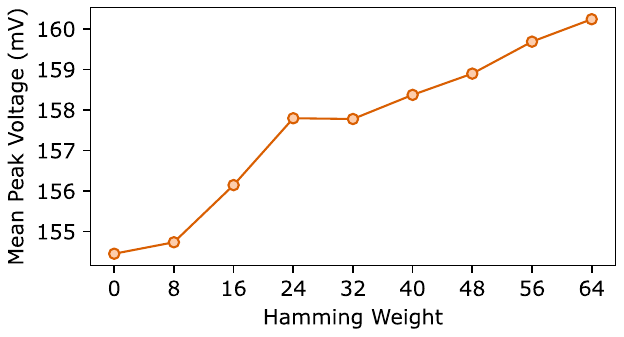}
        \caption{\texttt{HWT\_VALUE} = 32}
    \end{subfigure}
    \hfill
    \begin{subfigure}{0.48\columnwidth}
        \centering
        \includegraphics[width=\textwidth]{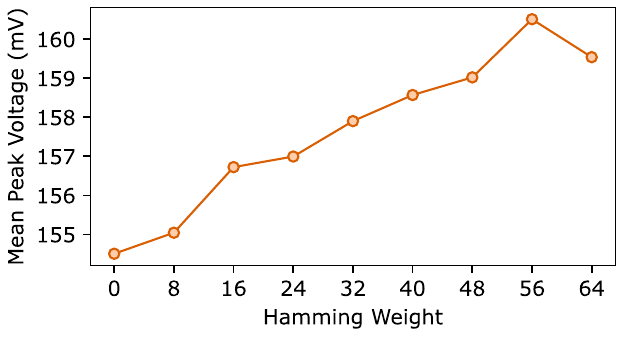}
        \caption{\texttt{HWT\_VALUE} = 48}
    \end{subfigure}

    \vspace{0.2cm} 
    
    \begin{subfigure}{0.48\columnwidth}
        \centering
        \includegraphics[width=\textwidth]{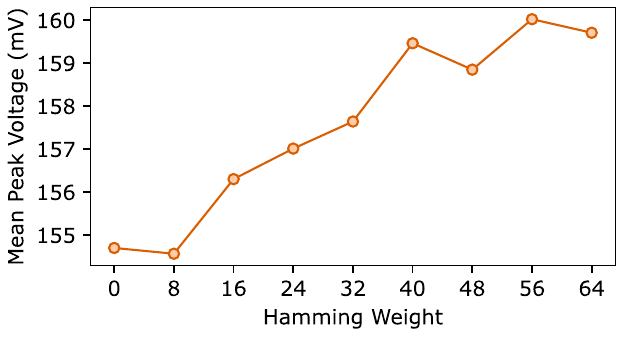}
        \caption{\texttt{HWT\_VALUE} = 64}
    \end{subfigure}
    \hfill
    \caption{Effect of the HWT of data processed prior to the context switch (\texttt{HWT\_VALUE\_LOOP}), reflecting residual power influence on the spike. The HWT of the data in the registers (\texttt{HWT\_VALUE}) is held constant while varying the HWT of the data being processed inside the loop. Each figure illustrates the impact of HWT of data processed before the context switch when the register HWTs are fixed at 0, 16, 32, 48, and 64.}
    \label{fig:residual_HW_results}
    \vspace{-0.35in}
\end{figure}

However, as shown in Figure~\ref{fig:residual_HW_results}, the HWT of data processed earlier also exhibits a strong correlation with the observed power signature, as expected due to the power-temperature interdependency. Thus, we validate Hypothesis~2, confirming that both the HWT of the data in the registers during the context switch and the HWT of the data processed prior to the context switch influence the observed power consumption.

\vspace{0.1in}
\subsection{Hamming Distance of the Processed Data}
\label{sec:residual_HD}

 In order to evaluate the effect of switching of data between \texttt{1 $\rightarrow$ 0} and \texttt{0 $\rightarrow$ 1} before a context switch, we focus on the HD of the processed data. We construct the following hypothesis to be evaluated during this experiment.

\begin{tcolorbox}[highlightbox]
Hypothesis 3: The effect of the Hamming distance of the data, which was processed earlier, has an effect on the \texttt{sleep}-induced power spike.
\end{tcolorbox}

To evaluate the hypothesis, we have created the program illustrated in Figure~\ref{fig:residual_HD}. The objective of this program is to isolate the effect of HD by controlling other variables, such as the HWT of individual states. To facilitate isolation, the program is designed with alternating groups of two \texttt{LEFT SHIFT} and \texttt{RIGHT SHIFT} operations. The reason for implementing \texttt{SHIFT} operations in pairs is based on the microarchitectural characteristics of the ARM Cortex-A72 processor, which features eight execution pipelines. Among these, two pipelines support integer ALU micro-operations, including \texttt{LEFT SHIFT (LSL)} and \texttt{RIGHT SHIFT (RSL)} instructions~\cite{arm2015cortexa72}.
 The \texttt{SHIFT} operation can maintain the HWT of the input and output data within a certain range while introducing an HD effect. We implemented the shift operation with the help of two input registers: one containing data with a fixed HWT, and the other containing the shift amount. The other general-purpose registers were also loaded with data of the same fixed HWT and were used as output registers for the \texttt{SHIFT} operations. Note that in each iteration, the same output registers are used for both \texttt{LEFT SHIFT} and \texttt{RIGHT SHIFT} operations. This design facilitates bit transitions within the registers beyond the ALU output, due to the alternating application of \texttt{LSL} and \texttt{RSL} instructions. We choose the fixed value as \texttt{0x00000ffffff00000} and define a variable named \texttt{SHIFT\_VALUE} that can take an arbitrary number of bits. After the first iteration in the loop, when the shift operation is executed, the $2 \times \texttt{SHIFT\_VALUE} $ number of bits immediately before the first `1' bit in the output register changes from 0 to 1. At the same time, the last $2 \times \texttt{SHIFT\_VALUE}$ number of `1' bits change from 1 to 0. This results in an HD of $4 \times \texttt{SHIFT\_VALUE}$ in the output register and ALU output. For example, if $\texttt{SHIFT\_VALUE} = 4$, after the first iteration in the loop, the output registers and ALU output contain \texttt{0x000000ffffff0000} (because of the RIGHT SHIFT), and when the LEFT SHIFT is performed (on \texttt{0x00000ffffff00000}) the red-colored 8 bits transition from 0 to 1, while the green-colored 8 bits transition from 1 to 0, leading to a total HD of 16:
\[
\texttt{0x0000\textcolor{red}{00}ffff\textcolor{green}{ff}0000} \rightarrow \texttt{0x0000\textcolor{red}{ff}ffff\textcolor{green}{00}0000}
\]
The subsequent \texttt{RIGHT SHIFT} operation performs a similar transition (from \texttt{0x0000ffffff000000} to \texttt{0x000000ffffff0000}), restoring the value in the output register to its initial state. 

\begin{figure}
    \centering
    \tiny
    \begin{subfigure}{0.48\columnwidth}
        \centering
        \includegraphics[width=\textwidth]{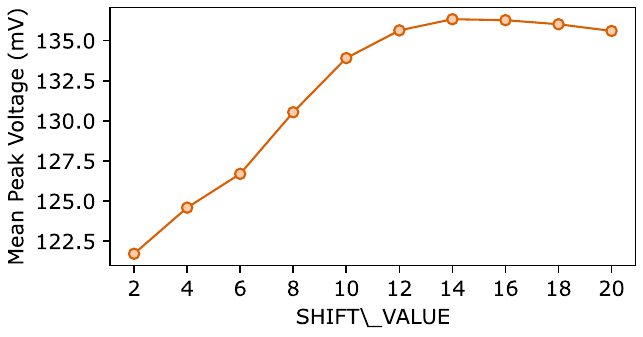}
        \caption{\texttt{HWT\_VALUE} = 0}
    \end{subfigure}
    \hfill
    \begin{subfigure}{0.48\columnwidth}
        \centering
        \includegraphics[width=\textwidth]{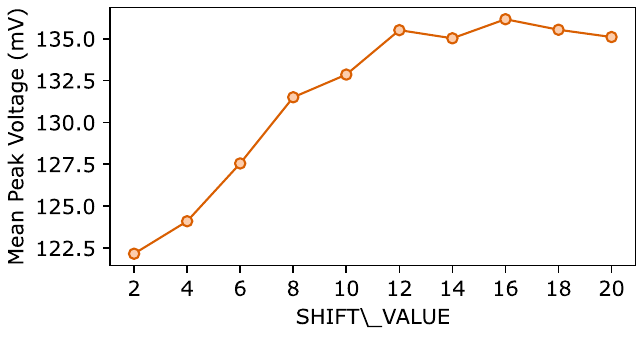}
        \caption{\texttt{HWT\_VALUE} = 16}
    \end{subfigure}
    
    \vspace{0.2cm} 
    
    \begin{subfigure}{0.48\columnwidth}
        \centering
        \includegraphics[width=\textwidth]{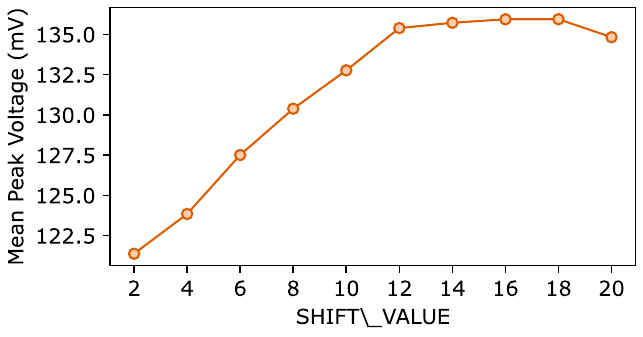}
        \caption{\texttt{HWT\_VALUE} = 32}
    \end{subfigure}
    \hfill
    \begin{subfigure}{0.48\columnwidth}
        \centering
        \includegraphics[width=\textwidth]{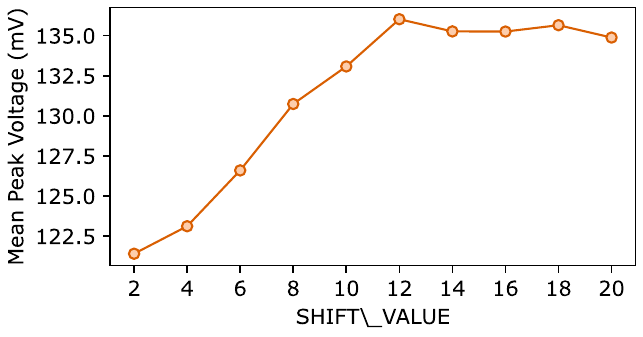}
        \caption{\texttt{HWT\_VALUE} = 48}
    \end{subfigure}

    \vspace{0.2cm} 
    
    \begin{subfigure}{0.48\columnwidth}
        \centering
        \includegraphics[width=\textwidth]{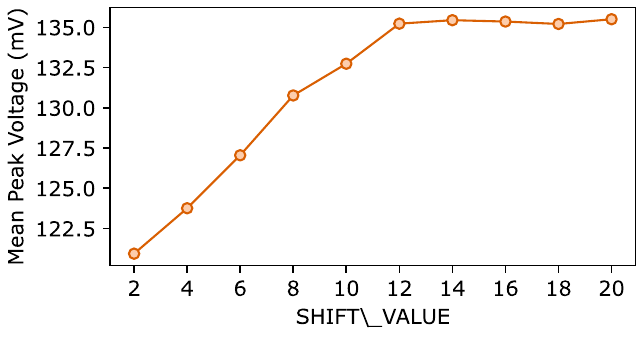}
        \caption{\texttt{HWT\_VALUE} = 64}
    \end{subfigure}
    \hfill
    \caption{The effect of varying HD of processed data prior to the context switch with \texttt{SHIFT\_VALUE} amount, while keeping the HWT of the data in the registers (\texttt{HWT\_VALUE}) constant. Different HD values were achieved by changing the shift amount. Since we use $0x00000ffffff00000$ as the fixed value, the HD is always constant ($4 \times \text{shift amount}$).}
    \label{fig:residual_HD_results}
    \vspace{-0.15in}
\end{figure}

As illustrated in Figure~\ref{fig:residual_HD_results}, the HD of the data processed prior to the context switch influences the power spike amplitude observed during the context switch. This result supports our hypothesis for this specific experiment as well as the explanation of the power-temperature feedback.

\begin{figure}[ht]
  \centering
  \scriptsize
  \begin{minipage}{0.9\columnwidth}
    \begin{center} 
    \begin{lstlisting}
// Read shift amount
X0 = SHIFT_VALUE 
// Value with fixed HWT
X1 = 0x00000ffffff00000 
// Number of loop iterations
X10 = LOOP_ITR
// Load all GPR with the same value
MOV X2, X1         
::
MOV X26, X1
const_loop: 
    // Perform Shift operations
    LSL X2, X1, X0 
    LSL X3, X1, X0
    ::
    RSL X2, X1, X0
    RSL X3, X1, X0
    ::
    B const_loop
end const_loop: 
    // Value with fixed HWT 
    X0 = HWT_VALUE 
    // Load all GPR with the same value
    MOV X1, X0         
    ::
    MOV X26, X0 
    // Make a system call for sleep
    ADR X0, timespec      
    MOV X1, XZR             
    MOV X8, #101            
    SVC #0   
    \end{lstlisting}
    \end{center}
  \end{minipage}
  \vspace{0.15in}
  \caption{Code for residual power HD experiment.}
  \label{fig:residual_HD}
  \vspace{-0.25in}
\end{figure}

\subsection{Number of Executed Instructions}
\label{sec:residual_num_inst}
To evaluate the impact of the number of instructions executed prior to a context switch on the \texttt{sleep}-induced power spike, we formulate the following hypothesis.

\begin{tcolorbox}[highlightbox]
Hypothesis 4: The number of executed instructions earlier has an effect on the \texttt{sleep}-induced power spike.
\end{tcolorbox}

To test this hypothesis, we conducted experiments using the setup described in Section~\ref{subsec:setup}, systematically varying the number of instructions executed before the context switch. This was accomplished by adjusting the loop iteration count (\texttt{LOOP\_ITR}) in the same programs used in our earlier experiments, as shown in Figures~\ref{fig:residual_HW} and~\ref{fig:residual_HD}.
In the experiment based on the program in Figure~\ref{fig:residual_HW}, we defined two scenarios: \textbf{Cold Task} and \textbf{Hot Task}. In the Cold Task, both \texttt{HWT\_VALUE\_LOOP} and \texttt{HWT\_VALUE} were configured to produce a HWT of 0. In the Hot Task, both were set to a HWT of 64. Similarly, in the experiment using the program from Figure~\ref{fig:residual_HD}, the Cold Task was configured with \texttt{SHIFT\_VALUE} set to 2 and \texttt{HWT\_VALUE} to a HWT of 0, whereas the Hot Task used a \texttt{SHIFT\_VALUE} of 20 and \texttt{HWT\_VALUE} set to a HWT of 64.

\begin{figure}
    \centering
    \tiny
    \begin{subfigure}{0.48\columnwidth}
        \centering
        \include{images/Trace507_510}
        \vspace{-0.25in}
        \caption{Power traces from Figure~\ref{fig:residual_HW} with two loop iteration counts.}
        \label{fig:residual_num_instructions_trace_HW}
    \end{subfigure}
    \hfill
    \begin{subfigure}{0.48\columnwidth}
        \centering
        \include{images/Trace407_410}
        \vspace{-0.25in}
        \caption{Power traces from Figure~\ref{fig:residual_HD} with two loop iteration counts.}
        \label{fig:residual_num_instructions_trace_HD}
    \end{subfigure}
    
    \vspace{0.2cm} 
    
    \begin{subfigure}{0.48\columnwidth}
        \centering
        \includegraphics[width=\textwidth]{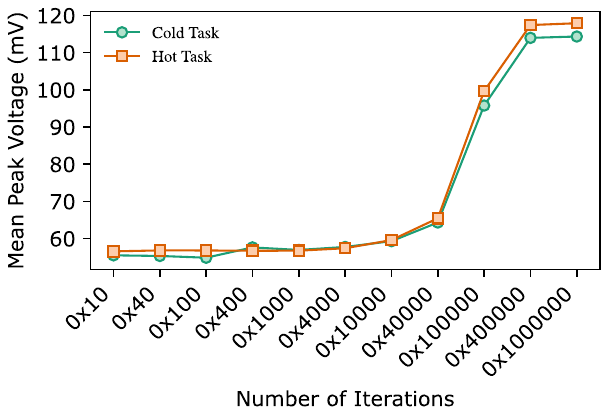}
        \caption{Effect of varying loop iterations in program~\ref{fig:residual_HW}. } 
        \label{fig:residual_num_instructions_HW}
    \end{subfigure}
    \hfill
    \begin{subfigure}{0.48\columnwidth}
    \centering
    \includegraphics[width=\textwidth]{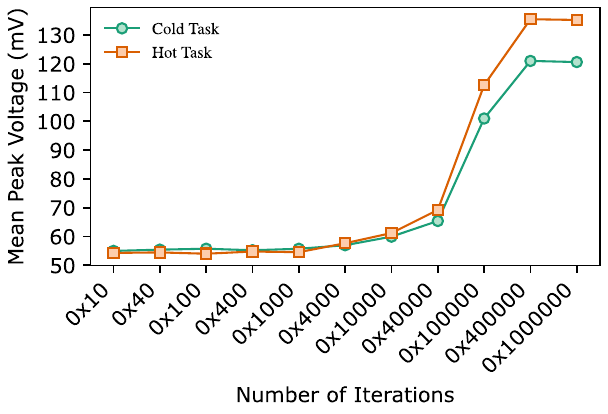}
        \caption{Effect of varying loop iterations in program~\ref{fig:residual_HD}.}
        \label{fig:residual_num_instructions_HD}
    \end{subfigure}

    \hfill
    \caption{Impact of loop iteration count on the \texttt{sleep}-induced power spike. At low iteration counts, the mean peak power remains low. As the iteration count increases, the mean peak power rises, reflecting the influence of accumulated instruction execution on residual power.}
    \label{fig:residual_num_instructions_results}
    \vspace{-0.15in}
\end{figure}

The results of this experiment are presented in Figure~\ref{fig:residual_num_instructions_results}. As shown in Figures~\ref{fig:residual_num_instructions_trace_HW} and~\ref{fig:residual_num_instructions_trace_HD}, a clear difference in spike amplitude emerges between programs with fewer loop iterations and those with more. This confirms that residual power is influenced by the number of executed instructions, specifically, an increase in instruction count amplifies the residual component of the \texttt{sleep}-induced power spike. Figures~\ref{fig:residual_num_instructions_HW} and~\ref{fig:residual_num_instructions_HD} further illustrate this effect by showing how varying the loop iteration count (\texttt{HWT\_VALUE\_LOOP}) impacts both Cold and Hot Tasks for the programs in Figures~\ref{fig:residual_HW} and~\ref{fig:residual_HD}. In both cases, the power spike amplitude increases with more iterations, and Hot Tasks consistently produce a higher mean peak power than Cold Tasks.


\subsection{Exploiting the \textit{SleepWalk} Power Model}

The presence of a residual power signature in the \texttt{sleep}-induced power spike significantly enhances the applicability and effectiveness of this novel side-channel. It enables the exploitation of both the data present in registers during the context switch via the power signature of the context-switch and information about previously executed programs and corresponding data via the residual power signature. We leverage this vulnerability to perform two distinct attacks, as demonstrated in the subsequent sections: a key recovery attack on SIKE (Section~\ref{sec:sike}) and a key recovery attack on AES (Section~\ref{sec:aes}).
Specifically, these two attacks explore two complementary strategies: a context switch power signature-dominant attack (Section~\ref{sec:sike_level_1}), where we insert or trigger context switches at precise points in the victim program to extract data in registers; and a residual power signature-dominant attack (Sections~\ref{sec:sike_level_2} and ~\ref{sec:aes}), where we repeatedly execute the same victim program to extract secret information based on accumulated power signatures.



\color{black}

\section{Key Recovery Attack on SIKE}
\label{sec:sike}

In this section, we perform a known-ciphertext proof-of-concept attack on the Supersingular Isogeny Key Encapsulation (SIKE)~\cite{campagna2019supersingular} implementation using the experimental setup outlined in Section~\ref{subsec:setup}.    
SIKE is a post-quantum cryptographic scheme designed for secure key exchange that utilize the properties of supersingular elliptic curves. It is based on the mathematical framework of isogeny graphs, which represent connections between elliptic curves through isogenies. SIKE builds upon the Supersingular Isogeny Diffie-Hellman (SIDH) protocol but incorporates additional mechanisms, such as the Fujisaki–Okamoto transformation, to provide stronger security guarantees. SIKE is particularly notable for its compact public keys and ciphertexts. Cryptographic operations in SIKE involve scalar multiplications, isogeny computations, and evaluations, making the protocol efficient but vulnerable to certain side-channel attacks. Although SIKE has been deprecated due to unrelated security concerns, the purpose of this attack is to demonstrate how to exploit cryptographic algorithms with input-dependent intermediate computation using  \texttt{sleep}-induced power spike.

For this experiment, we used the SIKE implementation available in the Interoperable Reusable Cryptographic Library (CIRCL)~\cite{armando2019introducing} of \textit{Cloudflare}.
Key extraction from SIKE focuses on exploiting zero-value intermediate states, which arise due to the mathematical structure and behavior of elliptic curve operations in SIKE~\cite{wang2022hertzbleed, de2022sike}. Adversaries can generate specific input ciphertexts, such that they produce specific computations during the victim's decapsulation process that produce anomalous zero values. These anomalies occur when the provided input ciphertexts force computations involving elliptic curve points, resulting in zero coordinates during the Montgomery three-point ladder. Let the secret key $k$ is represented as a binary value, and the goal of the adversary is to recover each bit sequentially. When the specially generated ciphertext targets the $i$-th bit of $k$, assuming the adversary knows up to the $(i-1)$-th bit of $k$, anomalous zero values are generated during the $(i+1)$-th round of the Montgomery ladder computation, if $k[i-1] \neq k[i]$, where $k[i-1]$ and $k[i]$ represent the $i-1$-th and $i$-th bits of $k$,  respectively. Once these anomalous zeros are introduced, the computation becomes stuck, propagating zeros through all subsequent intermediate values.
These zero values further influence subsequent computations, including isogeny evaluations and the j-invariant computation. The introduction of zero-value states results in consistent patterns in power consumption, as arithmetic operations involving zero inputs are computationally simpler. This propagation, along with noticeable changes in the HWT within the registers, creates a strong and detectable signature in the power spike. Based on the inherent weakness in the SIKE algorithm, we conduct two separate experiments as follows,

\begin{itemize}
    \item Proof-of-concept (PoC) attack on the SIKE implementation to evaluate feasibility, demonstrating a context-switch power signature-dominant attack.
    \item Attack on the original SIKE implementation in CIRCL, demonstrating a residual power signature-dominant attack.
\end{itemize}


\subsection{Proof-of-concept attack} 
\label{sec:sike_level_1}

\begin{figure}[htp]
\centering
\vspace{-0.1in}
\scriptsize
\includegraphics[width=0.4\textwidth]{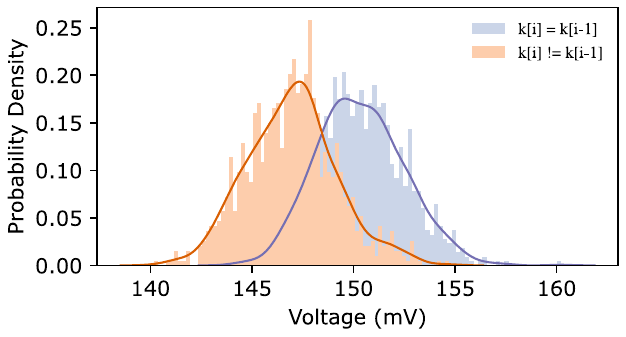}
\vspace{-0.1in}
\caption{Probability distribution of measured \texttt{sleep}-induced peak power amplitudes over the first ten bits of the secret key used in SIKE implementation.}
\label{fig:SIKE_distribution}
\vspace{-0.1in}
\end{figure}

For this experiment, we modified the CIRCL SIKE implementation by adding a \texttt{sleep} function call immediately after the Montgomery three-point ladder function, where zeros are generated for specific ciphertexts if the above condition is met. This modification enabled us to observe the ability to capture anomalous zeros using \texttt{sleep}-induced power spikes, characterizing this as a context switch power signature-dominant attack. Then, for a randomly selected key, we performed SIKE decapsulation with specially crafted ciphertexts. To evaluate the power signature difference for the conditions $k[i-1] \neq k[i]$ and $k[i-1] = k[i]$ for a target key bit $i$, we created two ciphertexts: one assuming $k[i-1] \neq k[i]$ and the other assuming $k[i-1] = k[i]$. For each key bit in the secret key $k$, we captured 1000 \texttt{sleep}-induced peak power values and measured the peak amplitude. Capturing 1000 traces took approximately 5 minutes. Figure~\ref{fig:SIKE_distribution} illustrates the distribution of measured peak amplitudes over the first ten bits of the secret key. The results confirm that there is a detectable difference in peak power amplitude between cases where an anomalous zero is triggered ($k[i-1] \neq k[i]$) and where it is not ($k[i-1] = k[i]$).

\begin{figure*}[htp]
\centering
\begin{subfigure}{\columnwidth}
    \centering
    \scriptsize
    \includegraphics[height=1.6in]{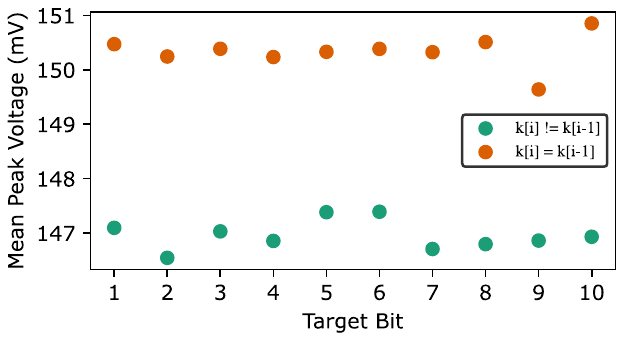}
    \vspace{-0.1in}
    \caption{Results of Level 1 SIKE Attack, where we invoked the \texttt{sleep} after the  Montgomery three-point ladder}
    \label{fig:sike_mean_value_inside_sleep}
\end{subfigure}
  \hfill
\begin{subfigure}{\columnwidth}
    \centering
    \scriptsize
    \includegraphics[height=1.6in]{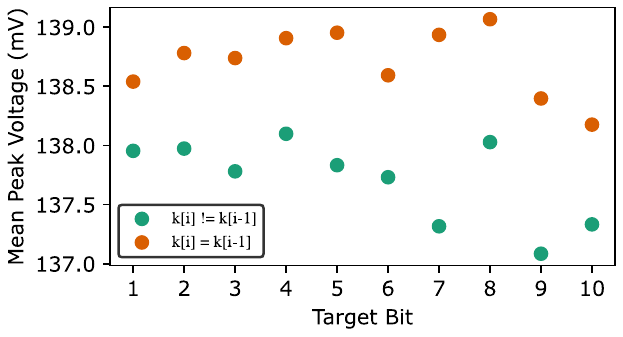}
    \vspace{-0.1in}
    \caption{Results of Level 2 SIKE Attack, where we invoked the \texttt{sleep} after the original SIKE decapsulation process}
    \label{fig:sike_mean_value_seperate_sleep}
\end{subfigure}
\caption{Mean values of the measured \texttt{sleep}-induced peak power amplitudes over the first ten bits of the SIKE secret key for both cases: when an anomalous zero is triggered ($k[i-1] \neq k[i]$; green colored data points) and when it is not ($k[i-1] = k[i]$; orange colored data points)}
\label{fig:sike_mean_values}
\vspace{-0.1in}
\end{figure*}

Next, we perform a complete key extraction attack with a threshold-based key bit identification approach. We plot the mean values of the measured peak amplitudes over the first ten bits of the secret key for both cases: when an anomalous zero is triggered ($k[i-1] \neq k[i]$) and when it is not ($k[i-1] = k[i]$). As shown in Figure~\ref{fig:sike_mean_value_inside_sleep}, there is a clear separation between the anomalous zero-triggered case ($k[i-1] \neq k[i]$) and the non-triggered case ($k[i-1] = k[i]$), indicating that a simple threshold-based classification is possible to perform the attack. 
The reason behind this distinct separation is that since we force a context switch immediately after the Montgomery three-point ladder operation, the registers are loaded with zeros in the anomalous zero-triggered case. Consequently, the \texttt{sleep}-induced power spike reflects a lower HWT as explained in Section~\ref{sec:leakage_model}. As shown in Figure~\ref{fig:sike_mean_value_inside_sleep}, when an anomalous zero is triggered, the mean value of the measured peak amplitudes is lower compared to the non-triggered case.

\subsection{SIKE Attack} 
\label{sec:sike_level_2}

In order to create a realistic attack scenario with a strict threat model, we use the original SIKE implementation in CIRCL. In a separate program, we execute the SIKE decapsulation function and call the \texttt{sleep} function immediately after the execution of the decapsulation function. This ensures that the SIKE implementation itself remains unmodified, while the attacker is utilizing a custom user-space program to invoke the decapsulation function. This attack does not benefit from scenarios where context switching occurs immediately after anomalous zeros appear in the registers, as seen in the previous PoC SIKE attack. This limitation arises because the context switch is deliberately triggered only at the end of the decapsulation process. However, the measured peak values still contain residual power signatures from the executed program and this residual power correlates with the HWT of the data processed during execution as explained in Section~\ref{sec:leakage_model}. As a result, even when the context switch is applied at the end, the peak power values still carry information useful for key recovery.

During the experiment, we found that a single decapsulation does not provide sufficient residual power to extract the key reliably. To enhance the amount of information captured in the peak values, we execute four iterations of decapsulation before forcing the context switch. 
After applying this technique, we analyze the mean values of the measured peak amplitudes over the first ten bits of the secret key for both cases: when an anomalous zero is triggered ($k[i-1] \neq k[i]$) and when it is not ($k[i-1] = k[i]$). Figure~\ref{fig:sike_mean_value_seperate_sleep} illustrates the results of this experiment. It can be observed that the results do not exhibit a consistent separation between the ($k[i-1] \neq k[i]$) and ($k[i-1] = k[i]$) cases, unlike the previous Level 1 SIKE attack shown in Figure~\ref{fig:sike_mean_value_inside_sleep}. For each bit position, the mean value for the ($k[i-1] \neq k[i]$) case is always lower than the mean value for the ($k[i-1] = k[i]$) case. This is caused by the residual power of the program retained within the peak power values. 

\begin{algorithm}
\small
\caption{Key Recovery Algorithm}
\begin{algorithmic}[1]
\State Recovered key: $K \gets 0$
\State Target bit position: $t$
\For{$t = 1$ to SIKE\_KEY\_LENGTH}
    \State $K[t] \gets \neg K[t-1]$
    \State $c \gets \text{generate\_ciphertext}(K, t)$
    \State collected\_traces $\gets$ perform\_attack($c, K$)
    \State peak\_values $\gets$ find\_peak(collected\_traces)
    \State mean\_value\_1 $\gets \text{mean}(\text{peak\_values})$

    \\
    \State $K[t] \gets K[t-1]$
    \State $c \gets \text{generate\_ciphertext}(K, t)$
    \State collected\_traces $\gets$ perform\_attack($c, K$)
    \State peak\_values $\gets$ find\_peak(collected\_traces)
    \State mean\_value\_2 $\gets \text{mean}(\text{peak\_values})$

    \\
    \If{$|$mean\_value\_1 $-$ mean\_value\_2$| < $SEP\_LIMIT}
        \State $K[t] \gets K[t-1]$
    \ElsIf{mean\_value\_2 $-$ mean\_value\_1 $\geq$ SEP\_LIMIT}
        \State $K[t] \gets \neg K[t-1]$
    \EndIf
\EndFor
\end{algorithmic}
\label{algo:sike_key_recovery_seperate_sleep}
\end{algorithm}

Unlike in the previous PoC SIKE attack, a threshold-based separation of two conditions is not feasible for this attack. Therefore, we employ Algorithm~\ref{algo:sike_key_recovery_seperate_sleep} to extract the key. We have to perform the attack iteratively while recovering the key bit by bit sequentially with the algorithm. First, we set the target key bit $K[t]$ as the flipped value of the previously recovered key bit. That is, if $K[t-1]$ is 1, then $K[t]$ is initially set to 0, and vice versa. We then generate the specific ciphertext $c$ using the function $generate\_ciphertext$. Next, we execute the decapsulation setup with amplification using the function $perform\_attack$ to collect the power traces and extract the peak values. The peak values are then identified using the function $find\_peak$, and the mean peak value, $mean\_peak\_value_1$, is computed using the function $mean$. If our guess for $K[t]$ is correct, the ciphertext $c$ will induce anomalous zeros (since $K[t] \neq K[t-1]$) during SIKE decapsulation, resulting in a lower mean peak power value. Conversely, if our guess is incorrect, decapsulation will not produce anomalous zeros, as the ciphertext is generated using an incorrect key bit value, meaning $K[t] = K[t-1]$. This results in a higher mean peak power value. Since we have only one measured mean peak power value, we repeat the procedure by setting $K[t] = K[t-1]$. If this is the correct guess, both cases will yield high mean peak power values in the same range. However, if this guess is incorrect, the first case (where $K[t]$ was flipped) will produce a lower mean value due to the presence of anomalous zeros, whereas the second case will yield a higher mean value. At the end of this process, these conditions are checked, and based on the results, the correct key bit $K[t]$ can be determined before proceeding to the next key bit. The threshold (\texttt{SEP\_LIMIT}) used to determine whether two mean values are significantly different or within the same range. The threshold is set by conducting multiple local experiments to find the minimum distance between cases where anomalous zeros are present and absent. By executing the attack described in Algorithm~\ref{algo:sike_key_recovery_seperate_sleep}, we were able to recover the full secret key.






\section{Key Recovery Attack on AES}
\label{sec:aes}

In this section, we perform our final key recovery attack on an AES-128 software implementation using \textit{SleepWalk} side-channel. For this attack, we follow the chosen plaintext approach utilizing the \texttt{sleep}-induced power spike. Unlike the SIKE implementation, which contains a set of anomalous zeros that can be generated using specific ciphertexts during the decapsulation process, no known attacks exploit this specific characteristic with the AES implementation. Additionally, standard AES-128 processes data at the byte level, restricting the possible HWT values in registers to a range of 0 to 8. In our attack, we target the final round secret key, extracting one byte of the key at a time. For this attack, we use the experimental setup outlined in Section~\ref{subsec:setup}.

\begin{figure}[htp]
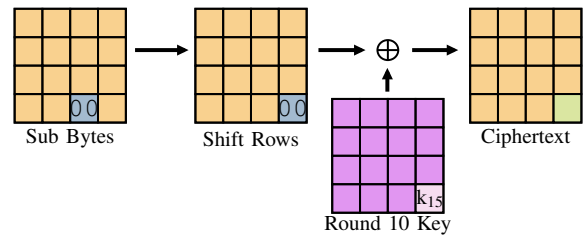

\centering
\include{images/AES_attack_method}
\vspace{-0.3in}
\caption{Attack method for AES-128. For plaintexts generated using Algorithm~\ref{algo:plaintxt_gen} with a successful final round key byte guess (e.g., $k_{15}$), the corresponding bytes in the output of the SubBytes and ShiftRows are \texttt{0x00} (represented by blue-colored boxes). For all other incorrect key guesses, the values in the blue-colored boxes differ from \texttt{0x00}.
}
\label{fig:AES_final_round}
\vspace{-0.1in}
\end{figure}

In an AES implementation, there are three main stages in the final round. As shown in Figure~\ref{fig:AES_final_round}, these stages are, 
\begin{enumerate}
    \item The SubBytes (S-box) operation.
    \item The ShiftRows operation.
    \item The XOR operation with the final round secret key.
\end{enumerate}
The final round is the ideal place to perform the attack since manipulating a value generated during the SubBytes (S-box) stage causes it to propagate to the ShiftRows stage. For example, when a byte becomes $0x00$ in the SubBytes stage, it propagates through the ShiftRows stage without modification, though its position may change. The corresponding byte in the final ciphertext will then be equal to the round 10 key byte at the same position as in the ShiftRows stage. Additionally, having the same value in both the SubBytes and ShiftRows stages amplifies the effect of that value's HWT on the residual power, minimizing algorithmic noise.
We conduct three separate attacks with AES as follows,

\begin{itemize}
    \item \textbf{Level 1 AES Attack}: Reconstruct the SubBytes output.
    \item \textbf{Level 2 AES Attack}: Detecting byte-level changes in the registers.
    \item \textbf{Level 3 AES Attack}: Extracting the final round key.
\end{itemize}

\begin{algorithm}
\caption{Plaintext Generator Algorithm}
\label{algo:plaintxt_gen}
\small
\begin{algorithmic}[1]
\State \textbf{Inputs}: Target key byte $(T)$
\State \textbf{Output}: Plaintext $(P)$
\State Ciphertext: $C \gets Rand()$
\State Key byte guess: $k$
\For{$k = 0$ to $255$}
    \State $C[T] \gets k$
    \State $P \gets   Decrypt(C) $
    
\EndFor
\end{algorithmic}
\end{algorithm}


\vspace{-0.15in}
\subsection{Level 1 AES Attack}

In order to perform a successful attack in real-world AES implementations, we have to address two key challenges: 1) how to manipulate a target byte in the final S-box stage and 2) how to distinguish single-byte HWT changes using \texttt{sleep}-induced power spikes. In order to solve the first challenge, we use Algorithm~\ref{algo:plaintxt_gen}. It effectively generates plaintexts such that they produce different HWT values at the targeted byte location $T$ in the final round ShiftRows stage of the AES implementation, for a given key byte guess of $k$. As shown in the algorithm, the attacker provides $T$ as input. For example, let us assume the target key byte is 15 ($T = 15$). For each target byte, we start with a random 16-byte value (line 3). Next, we replace the target byte with the key guess ($k$) (line 6). Finally, the plaintext is generated by decrypting the corresponding ciphertext (line 7). This algorithm guarantees that when the key guess ($k$) is correct, the byte at location $T$ in the final round ShiftRows stage becomes \texttt{0x00}, resulting in HWT 0. In all other cases, the byte at location $T$ in the final round ShiftRows stage will have a value with an HWT greater than 0. Therefore, by identifying the HWT 0 case from the rest of the $k$ guesses, we can recover the round 10 key byte.
Following this approach, we derive three plaintexts such that four SubBytes outputs in the final round are either \texttt{0x00}, \texttt{0xFF}, or random, creating three scenarios where the four values have HWT 0, HWT 8, or a random HWT. The remaining 12 bytes remain unchanged across the three scenarios. 

To strengthen the residual power effect, we executed 150,000 encryptions before invoking \texttt{sleep} to induce the power spike. This attack was conducted on the experimental setup outlined in Section~\ref{subsec:setup}. For each plaintext, we captured 10,000 traces, totaling 30,000 traces per test set, and we conducted two tests. Capturing 10,000 traces took approximately 6 hours. We then measured the peak value of each trace and calculated the mean value for each scenario. 
Next, we calculated the mean difference of the observed peak values between the random HWT and HWT 0 ({Random 1 – Four All 0s}), and between the random HWT and HWT 8 ({Random 1 – Four All 1s}) in the first test set, as illustrated in the first row of Table~\ref{tab:four_subbyte_comparision}. We repeated the same calculation for the second test set as well as shown in the second row of the Table~\ref{tab:four_subbyte_comparision}. By analysing both mean difference values, we observe that the scenario with four \texttt{0x00} values exhibits the highest difference, making it the easiest to distinguish from \texttt{0xFF} value. This confirms that the \texttt{sleep}-induced power spike can be used as a side-channel to recover the AES-128 final round key. 

\begin{table}[]
\caption{Comparison of average voltage difference for SubBytes outputs with random HWT, HWT 0, and HWT 8. The right two columns show voltage differences between random outputs and \texttt{0x00}, and between random outputs and \texttt{0xFF}. Rows correspond to the two test sets.}
\label{tab:four_subbyte_comparision}


\begin{tabular}{|c|c|c|}
\hline
\textbf{S-box Output} & \textbf{Four all 0s (\texttt{0x00})} & \textbf{Four all 1s (\texttt{0xFF})} \\ \hline
\text{Random 1} & \color[HTML]{00B050} 0.380544 mV & 0.189312 mV \\ \hline
\text{Random 2} & \color[HTML]{00B050} 0.345312 mV & 0.113056 mV \\ \hline
\end{tabular}
\vspace{-0.15in}
\end{table}

\subsection{Level 2 AES Attack}

To evaluate the fine granularity of \texttt{sleep}-induced power spike side-channel, we performed the level 2 attack on the AES implementation. The goal of this attack is to evaluate the detectability of a single-byte change in the final round key as shown in Figure~\ref{fig:AES_final_round}. Algorithm~\ref{algo:plaintxt_gen} with $T$ = 15 was used to evaluate the ability to differentiate single-byte HWT changes using \texttt{sleep}-induced power spike values. We conducted the experiment using three different secret keys, collecting 5,000 traces, which took approximately 3 hours for each guessed HWT of the last key byte. We perform encryption using generated plaintexts, and to amplify the residual power effect, we executed 150,000 encryptions before invoking the \texttt{sleep} function to collect the peak power spike values.
 
\begin{figure}[htp]
\centering
\vspace{-0.1in}
\scriptsize
\includegraphics[width=0.45\textwidth]{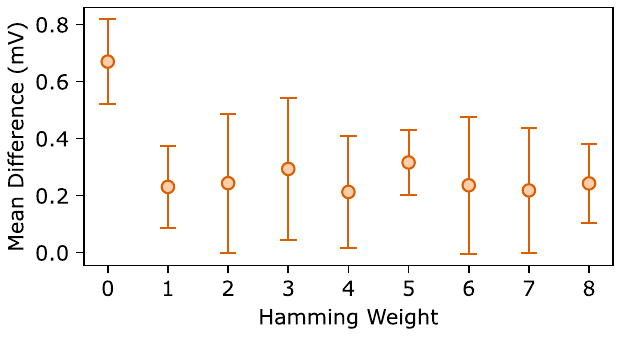}
\vspace{-0.1in}
\caption{Mean difference between the \texttt{sleep}-induced peak power of a given HWT and other HWT cases, over three different secret keys, with respect to varying HWT values in the last byte of the round 10 key.}
\label{fig:AES_HW_last_byte}
\vspace{-0.2in}
\end{figure}

Figure~\ref{fig:AES_HW_last_byte} presents the mean difference between the \texttt{sleep}-induced peak power of a given HWT and other HWT cases, evaluated over three different secret keys, with respect to varying HWT values in the last byte of the final round key. This means that the data point corresponding to HWT = 0 represents the mean difference in peak power values between the mean peak power of HWT = 0 and the mean peak powers of HWT = 1, 2, 3, 4, 5, 6, 7, and 8. We observe that HWT 0 exhibits the highest mean difference compared to other HWT values, making it detectable through its distinct peak power difference. This characteristic enables the classification of the HWT 0 case from the rest, allowing the key guess $k$ corresponding to HWT 0 to directly reveal the actual final round key byte.
Based on the results, we can conclude that it is possible to differentiate HWT 0 from the rest using \texttt{sleep}-induced power spike with single-byte resolution.

\subsection{Level 3 AES Attack}

Building on Level 1 and Level 2 attack results on AES, we create the Level 3 AES attack to evaluate the effectiveness of extracting the full final round key. We conducted the same experiment as in Level~2, varying the target byte $T$ in Algorithm~\ref{algo:plaintxt_gen} one at a time. For each target byte, we collected 5,000 traces over a period of approximately 3 hours and measured the mean peak power value.

\begin{figure}[htbp]
    \centering

    \begin{subfigure}{\linewidth}
        \centering
        \includegraphics[width=\linewidth]{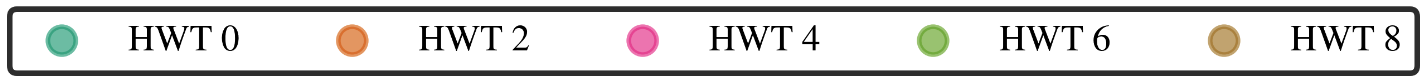}
    \end{subfigure}
    
    \begin{subfigure}{0.48\columnwidth}
        \centering
        \includegraphics[width=\linewidth]{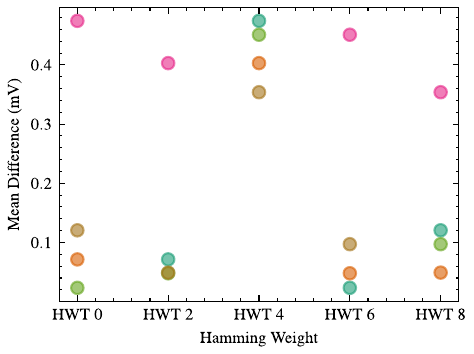}
        \vspace{-0.2in}
        \caption{Byte 15}
    \end{subfigure}
    \hfill
    \begin{subfigure}{0.48\columnwidth}
        \centering
        \includegraphics[width=\linewidth]{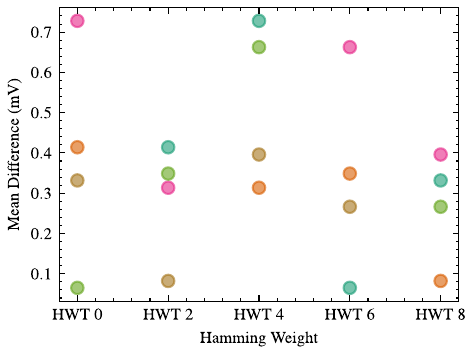}
        \vspace{-0.2in}
        \caption{Byte 13}
    \end{subfigure}
    
    
    \begin{subfigure}{0.48\columnwidth}
        \centering
        \includegraphics[width=\linewidth]{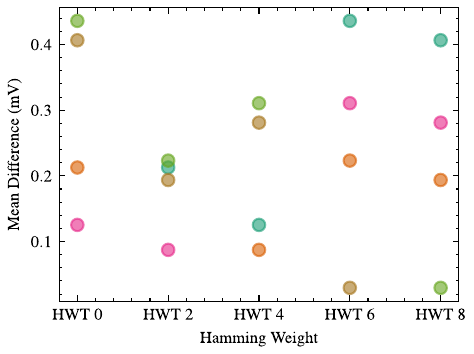}
        \vspace{-0.2in}
        \caption{Byte 3}
    \end{subfigure}
    \hfill
    \begin{subfigure}{0.48\columnwidth}
        \centering
        \includegraphics[width=\linewidth]{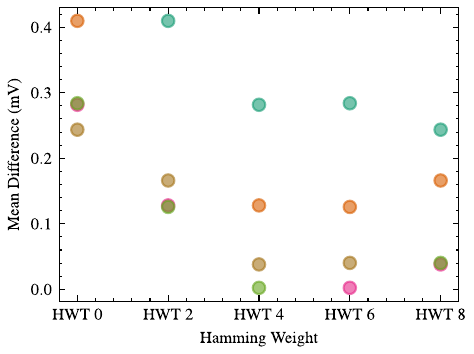}
        \vspace{-0.2in}
        \caption{Byte 1}
    \end{subfigure}
    
    \vspace{-0.05in}
    \caption{Differences between the mean \textit{sleep}-induced peak power of a given HWT and other HWT cases. Each colored data point represents the difference between the HWT indicated on the x-axis label and the HWT represented by the color.
}
    \label{fig:aes_final_round_key_bytes}
    \vspace{-0.15in}
\end{figure}

Figure~\ref{fig:aes_final_round_key_bytes} presents the differences between the mean \texttt{sleep}-induced peak power of a given HWT and other HWT cases, with respect to varying HWT values at byte locations 15, 13, 3, and 1 of the final round key. We observe that for each target byte, the difference between HWT 0 and another HWT produces the highest peak power variations, making these two cases the extreme ends of the peak power distribution for a target byte across different HWT values. 
For example, in target byte 15, the highest peak power difference is observed between HWT 0 and HWT 4, while in target byte 3, it occurs between HWT 0 and HWT 6. Since HWT 0 consistently represents the lower end of the power distribution for each target byte, the attacker can easily identify the peak power corresponding to HWT 0 and, consequently, determine the correct key byte values. 
Based on these results, we successfully recovered 10 out of 16 bytes of the final round key using \textit{SleepWalk}, demonstrating that a single power trace point can reduce the key recovery complexity from $2^{128}$ to $2^{48}$.

\section{Conclusion}
\label{sec:conclusion}

In this paper, we introduced \textit{SleepWalk}, a novel power side-channel vulnerability that exploits the \texttt{sleep}-induced power spike to perform cryptographic key recovery attacks on off-the-shelf hardware. We demonstrated that the power spike generated during the initialization of the \texttt{sleep} function carries the signature of context switching power consumption and residual power consumption from the previously executed computations. We explored a power model for \texttt{sleep}-induced spike, and established its correlation with data in registers and processed data before the context switch. We demonstrated two attacks exploiting this vulnerability. First, we successfully extracted the full secret key from a constant-time SIKE implementation using a novel single-point side-channel, eliminating the need for trace alignment, external triggers, or advanced preprocessing techniques. Finally, we developed a proof-of-concept chosen-plaintext attack on AES-128, proving that the \texttt{sleep}-induced power spike can be used to extract the final round key at byte granularity. 


\ifCLASSOPTIONcompsoc
  \section*{Code Availability and Disclosure}
\else
  \section*{Code Availability and Disclosure}
\fi

\textit{SleepWalk} framework with case studies can be downloaded from https://anonymous.4open.science/r/SleepWalk/ repository. 
We disclosed our findings, including the PoC code, to Broadcom and ARM in January 2025.


\vspace{-0.7in}
\begin{IEEEbiography}
[{\includegraphics[width=1in,clip, keepaspectratio]{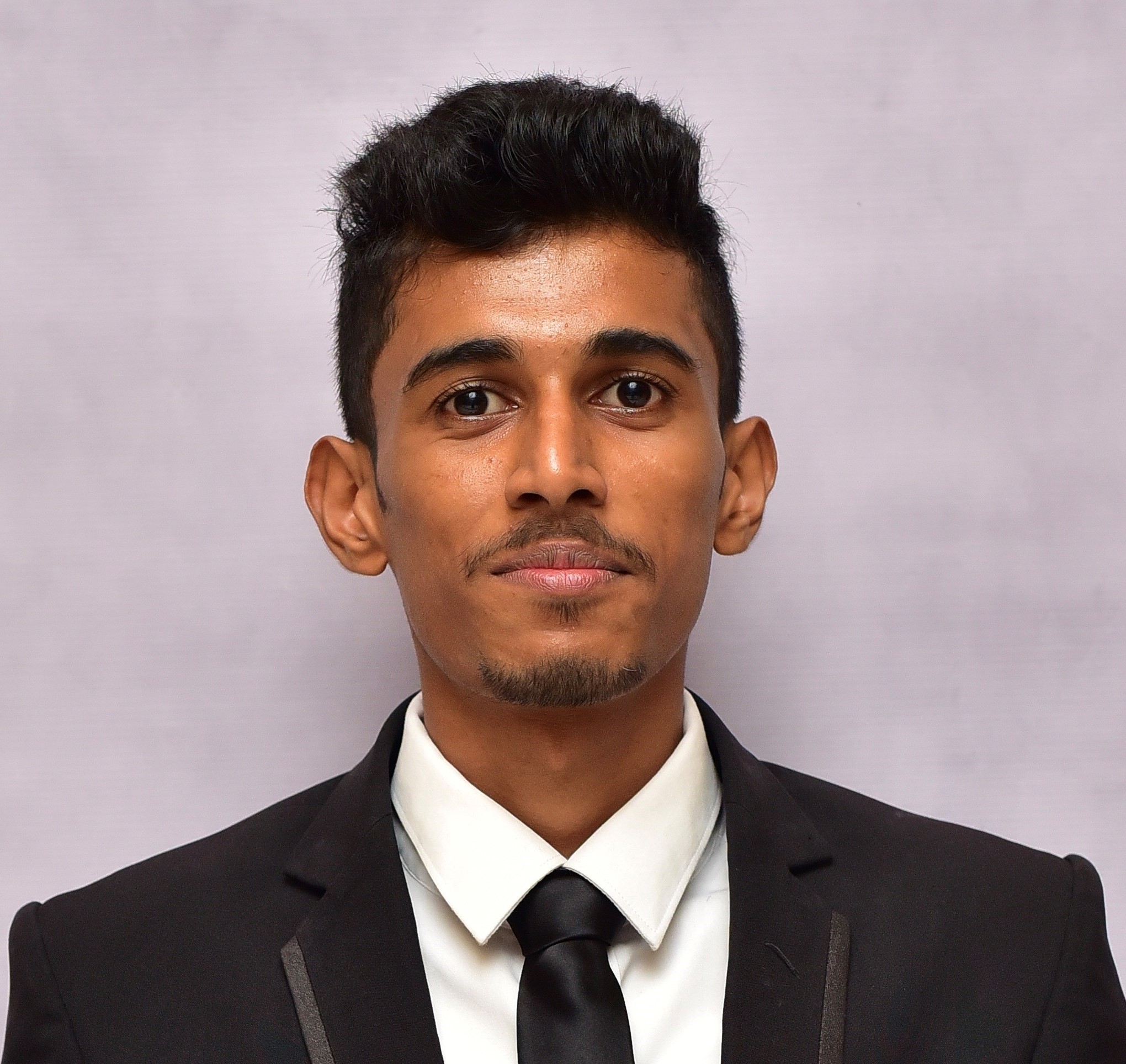}}]{Sahan Sanjaya} is a second-year Ph.D student in the Department of Computer \& Information Science \& Engineering at the University of Florida. In 2022, he completed his B.Sc. in the Department of Electronic and Telecommunication Engineering at the University of Moratuwa, Sri Lanka. His research interests encompass side-channel attacks, hardware security, pre-silicon validation, and post-silicon validation.
\end{IEEEbiography}

\vspace{-0.8 in}
\begin{IEEEbiography}[{\includegraphics[width=1in,clip, keepaspectratio]{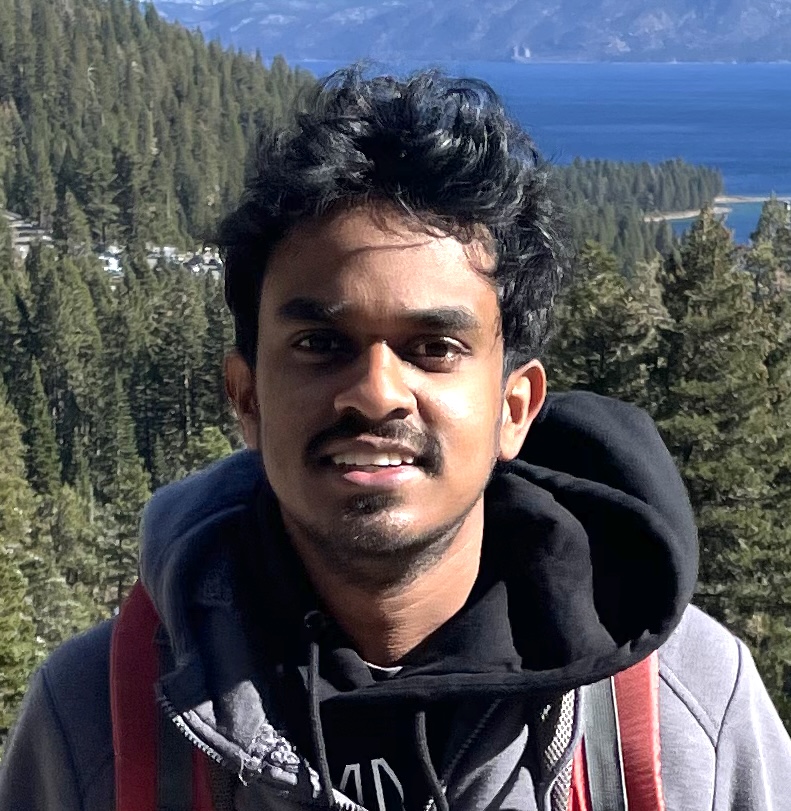}}]{Aruna Jayasena} is a Ph.D student in the Department of Computer \& Information Science \& Engineering at the University of Florida. He received his B.S. in the Department of Computer Science and Engineering at the University of Moratuwa, Sri Lanka, in 2019. His research focuses on systems security, hardware-firmware co-validation, test generation, trusted execution, side-channel analysis, and system-on-chip debug.
\end{IEEEbiography}

\vspace{-0.8 in}
\begin{IEEEbiography}[{\includegraphics[width=1in,clip,keepaspectratio]{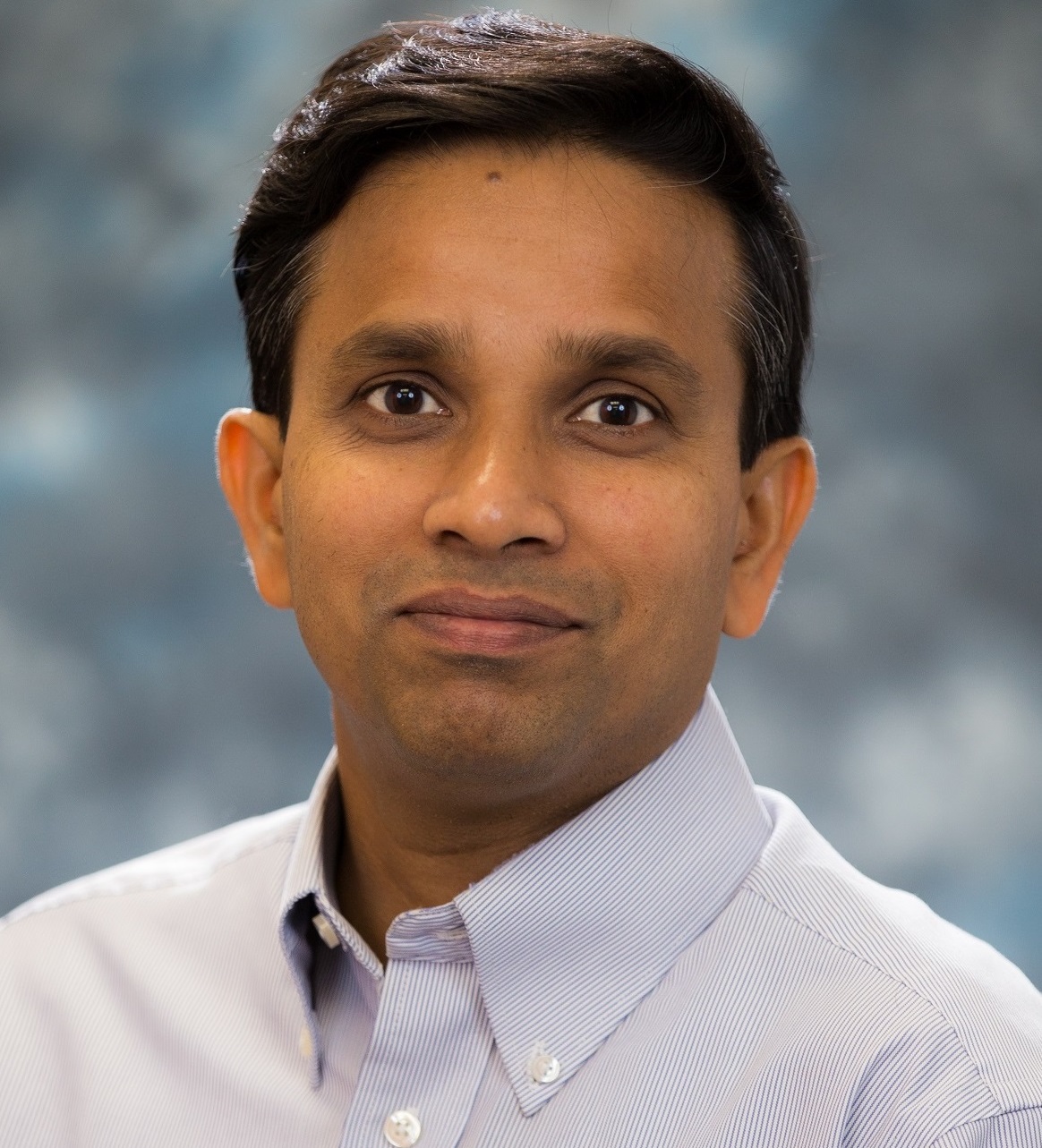}}]{Prabhat Mishra}
is a Professor in the Department of Computer and Information Science and Engineering at the University of Florida. 
His research interests include embedded and cyber-physical systems, hardware security and trust, and energy-aware computing. He currently serves as an Associate Editor of ACM Transactions on Design Automation of Electronic Systems and ACM Transactions on Embedded Computing Systems. He is an IEEE Fellow, an AAAS Fellow, and an ACM Distinguished Scientist.
\end{IEEEbiography}



\begin{thebibliography}{10}
\providecommand{\url}[1]{#1}
\csname url@samestyle\endcsname
\providecommand{\newblock}{\relax}
\providecommand{\bibinfo}[2]{#2}
\providecommand{\BIBentrySTDinterwordspacing}{\spaceskip=0pt\relax}
\providecommand{\BIBentryALTinterwordstretchfactor}{4}
\providecommand{\BIBentryALTinterwordspacing}{\spaceskip=\fontdimen2\font plus
\BIBentryALTinterwordstretchfactor\fontdimen3\font minus \fontdimen4\font\relax}
\providecommand{\BIBforeignlanguage}[2]{{%
\expandafter\ifx\csname l@#1\endcsname\relax
\typeout{** WARNING: IEEEtran.bst: No hyphenation pattern has been}%
\typeout{** loaded for the language `#1'. Using the pattern for}%
\typeout{** the default language instead.}%
\else
\language=\csname l@#1\endcsname
\fi
#2}}
\providecommand{\BIBdecl}{\relax}
\BIBdecl

\bibitem{li2007quantifying}
C.~Li, C.~Ding, and K.~Shen, ``Quantifying the cost of context switch,'' in \emph{Proceedings of the 2007 workshop on Experimental computer science}, 2007, pp. 2--es.

\bibitem{so2lecture3}
\BIBentryALTinterwordspacing
{Linux Kernel Labs}. (2023) So2 lecture 3: Processes. Accessed: 2025-03-21. [Online]. Available: \url{https://linux-kernel-labs.github.io/refs/heads/master/so2/lec3-processes.html}
\BIBentrySTDinterwordspacing

\bibitem{jayasena2024evilcs}
A.~Jayasena, R.~Bachmann, and P.~Mishra, ``Evilcs: An evaluation of information leakage through context switching on security enclaves,'' in \emph{2024 Design, Automation \& Test in Europe Conference \& Exhibition (DATE)}.\hskip 1em plus 0.5em minus 0.4em\relax IEEE, 2024, pp. 1--6.

\bibitem{kocher1999differential}
P.~Kocher, J.~Jaffe, and B.~Jun, ``Differential power analysis,'' in \emph{International Cryptology Conference}.\hskip 1em plus 0.5em minus 0.4em\relax Springer, 1999, pp. 388--397.

\bibitem{zhu2025controlled}
Y.~Zhu, B.~Chen, Z.~N. Zhao, and C.~W. Fletcher, ``Controlled preemption: Amplifying side-channel attacks from userspace,'' in \emph{ACM International Conference on Architectural Support for Programming Languages and Operating Systems}, 2025, pp. 162--177.

\bibitem{armando2019introducing}
\BIBentryALTinterwordspacing
A.~Faz-Hernandez and K.~Kwiatkowski. (2019) Introducing circl: An advanced cryptographic library. Cloudflare. Accessed: 2025-01-10. [Online]. Available: \url{https://github.com/cloudflare/circl}
\BIBentrySTDinterwordspacing

\bibitem{dworkin2001advanced}
M.~J. Dworkin, E.~Barker, J.~R. Nechvatal, J.~Foti, L.~E. Bassham, E.~Roback, J.~F. Dray~Jr \emph{et~al.}, ``Advanced encryption standard (aes),'' 2001.

\bibitem{gamaarachchi2018power}
H.~Gamaarachchi and H.~Ganegoda, ``Power analysis based side channel attack,'' 01 2018.

\bibitem{guillaume2022virtual}
J.~Guillaume, M.~Pelcat, A.~Nafkha, and R.~Salvador, ``Virtual triggering: a technique to segment cryptographic processes in side-channel traces,'' in \emph{2022 IEEE Workshop on Signal Processing Systems (SiPS)}.\hskip 1em plus 0.5em minus 0.4em\relax IEEE, 2022, pp. 1--6.

\bibitem{moos2019static}
T.~Moos, A.~Moradi, and B.~Richter, ``Static power side-channel analysis—an investigation of measurement factors,'' \emph{IEEE Transactions on Very Large Scale Integration (VLSI) Systems}, vol.~28, no.~2, pp. 376--389, 2019.

\bibitem{gu2023trace}
S.~Gu, Z.~Luo, Y.~Chu, Y.~Xu, Y.~Jiang, and J.~Guo, ``Trace alignment preprocessing in side-channel analysis using the adaptive filter,'' \emph{IEEE Transactions on Information Forensics and Security}, vol.~18, pp. 5580--5591, 2023.

\bibitem{kocher1996timing}
P.~C. Kocher, ``Timing attacks on implementations of diffie-hellman, rsa, dss, and other systems,'' in \emph{International Cryptology Conference}.\hskip 1em plus 0.5em minus 0.4em\relax Springer, 1996, pp. 104--113.

\bibitem{ahmadi2023shield}
M.~M. Ahmadi, F.~Khalid, R.~Vaidya, F.~Kriebel, A.~Steininger, and M.~Shafique, ``Shield: An adaptive and lightweight defense against the remote power side-channel attacks on multi-tenant fpgas,'' \emph{arXiv preprint arXiv:2303.06486}, 2023.

\bibitem{aysu2018binary}
A.~Aysu, M.~Orshansky, and M.~Tiwari, ``Binary ring-lwe hardware with power side-channel countermeasures,'' in \emph{2018 Design, Automation \& Test in Europe Conference \& Exhibition (DATE)}.\hskip 1em plus 0.5em minus 0.4em\relax IEEE, 2018, pp. 1253--1258.

\bibitem{brier2004correlation}
E.~Brier, C.~Clavier, and F.~Olivier, ``Correlation power analysis with a leakage model,'' in \emph{International Workshop on Cryptographic Hardware and Embedded Systems (CHES)}, 2004, pp. 16--29.

\bibitem{benhadjyoussef2021power}
N.~Benhadjyoussef, M.~Karmani, and M.~Machhout, ``Power-based side channel analysis and fault injection: Hacking techniques and combined countermeasure,'' \emph{International Journal of Advanced Computer Science and Applications}, vol.~12, no.~5, 2021.

\bibitem{gao2024deeptheft}
Y.~Gao, H.~Qiu, Z.~Zhang, B.~Wang, H.~Ma, A.~Abuadbba, M.~Xue, A.~Fu, and S.~Nepal, ``Deeptheft: Stealing dnn model architectures through power side channel,'' in \emph{2024 IEEE Symposium on Security and Privacy (SP)}.\hskip 1em plus 0.5em minus 0.4em\relax IEEE, 2024, pp. 3311--3326.

\bibitem{ahmed2023deep}
A.~A. Ahmed, R.~A. Salim, and M.~K. Hasan, ``Deep learning method for power side-channel analysis on chip leakages,'' \emph{Elektronika Ir Elektrotechnika}, vol.~29, no.~6, pp. 50--57, 2023.

\bibitem{yang2016inferring}
Q.~Yang, P.~Gasti, G.~Zhou, A.~Farajidavar, and K.~S. Balagani, ``On inferring browsing activity on smartphones via usb power analysis side-channel,'' \emph{IEEE Transactions on Information Forensics and Security}, vol.~12, no.~5, pp. 1056--1066, 2016.

\bibitem{matovu2020defensive}
R.~Matovu, A.~Serwadda, A.~V. Bilbao, and I.~Griswold-Steiner, ``Defensive charging: Mitigating power side-channel attacks on charging smartphones,'' in \emph{Proceedings of the Tenth ACM Conference on Data and Application Security and Privacy}, 2020, pp. 179--190.

\bibitem{chen2017powerful}
Y.~Chen, X.~Jin, J.~Sun, R.~Zhang, and Y.~Zhang, ``Powerful: Mobile app fingerprinting via power analysis,'' in \emph{IEEE Conference on Computer Communications}.\hskip 1em plus 0.5em minus 0.4em\relax IEEE, 2017, pp. 1--9.

\bibitem{lipp2021platypus}
M.~Lipp, A.~Kogler, D.~Oswald, M.~Schwarz, C.~Easdon, C.~Canella, and D.~Gruss, ``Platypus: Software-based power side-channel attacks on x86,'' in \emph{2021 IEEE Symposium on Security and Privacy (SP)}.\hskip 1em plus 0.5em minus 0.4em\relax IEEE, 2021, pp. 355--371.

\bibitem{zhang2021red}
Z.~Zhang, S.~Liang, F.~Yao, and X.~Gao, ``Red alert for power leakage: Exploiting intel rapl-induced side channels,'' in \emph{Proceedings of the 2021 ACM Asia Conference on Computer and Communications Security}, 2021, pp. 162--175.

\bibitem{wang2022hertzbleed}
Y.~Wang, R.~Paccagnella, E.~T. He, H.~Shacham, C.~W. Fletcher, and D.~Kohlbrenner, ``Hertzbleed: Turning power $\{$Side-Channel$\}$ attacks into remote timing attacks on x86,'' in \emph{USENIX Security Symposium}, 2022, pp. 679--697.

\bibitem{moos2017static}
T.~Moos, A.~Moradi, and B.~Richter, ``Static power side-channel analysis of a threshold implementation prototype chip,'' in \emph{Design, Automation \& Test in Europe Conference \& Exhibition (DATE), 2017}.\hskip 1em plus 0.5em minus 0.4em\relax IEEE, 2017, pp. 1324--1329.

\bibitem{sanjaya2025information}
S.~Sanjaya, A.~Jayasena, and P.~Mishra, ``Information leakage through physical layer supply voltage coupling vulnerability,'' \emph{IEEE Transactions on Very Large Scale Integration (VLSI) Systems}, 2025.

\bibitem{zhang2021psc}
T.~Zhang, J.~Park, M.~Tehranipoor, and F.~Farahmandi, ``{PSC-TG: RTL} power side-channel leakage assessment with test pattern generation,'' in \emph{ACM/IEEE Design Automation Conference (DAC)}, 2021, pp. 709--714.

\bibitem{pundir2022power}
N.~Pundir, J.~Park, F.~Farahmandi, and M.~Tehranipoor, ``Power side-channel leakage assessment framework at register-transfer level,'' \emph{IEEE Transactions on Very Large Scale Integration (VLSI) Systems}, 2022.

\bibitem{jayasena2025ciseleaks}
A.~Jayasena, R.~Bachmann, and P.~Mishra, ``Ciseleaks: Information leakage assessment of cryptographic instruction set extension prototypes,'' \emph{IEEE Transactions on Information Forensics and Security}, 2025.

\bibitem{he2019rtl}
M.~He, J.~Park, A.~Nahiyan, A.~Vassilev, Y.~Jin, and M.~Tehranipoor, ``{RTL-PSC}: Automated power side-channel leakage assessment at register-transfer level,'' in \emph{IEEE VLSI Test Symposium (VTS)}, 2019, pp. 1--6.

\bibitem{jayasena2023tvla}
A.~Jayasena, E.~Andrews, and P.~Mishra, ``{Test Vector Leakage Assessment on Hardware Implementations of Asymmetric Cryptography Algorithms},'' \emph{IEEE Transactions on Very Large Scale Integration (VLSI) Systems}, 2023.

\bibitem{kogler2023collide}
A.~Kogler, J.~Juffinger, L.~Giner, L.~Gerlach, M.~Schwarzl, M.~Schwarz, D.~Gruss, and S.~Mangard, ``$\{$Collide+ Power$\}$: Leaking inaccessible data with software-based power side channels,'' in \emph{32nd USENIX Security Symposium (USENIX Security 23)}, 2023, pp. 7285--7302.

\bibitem{lipp2022amd}
M.~Lipp, D.~Gruss, and M.~Schwarz, ``$\{$AMD$\}$ prefetch attacks through power and time,'' in \emph{31st USENIX Security Symposium (USENIX Security 22)}, 2022, pp. 643--660.

\bibitem{Keysight}
\BIBentryALTinterwordspacing
Keysight, ``Keysight dsox1102g user manual,'' 2023. [Online]. Available: \url{https://www.keysight.com/th/en/assets/7018-05520/data-sheets/5992-1965.pdf}
\BIBentrySTDinterwordspacing

\bibitem{linuxmanpage}
\BIBentryALTinterwordspacing
M.~Kerrisk. (2025) Linux manual page. man7.org. [Online]. Available: \url{https://man7.org/linux/man-pages}
\BIBentrySTDinterwordspacing

\bibitem{king2017stress}
C.~I. King, ``Stress-ng,'' \emph{URL: https://github.com/ColinIanKing/stress-ng}, vol.~39, 2017.

\bibitem{messerges2000using}
T.~S. Messerges, ``Using second-order power analysis to attack dpa resistant software,'' in \emph{International Workshop on Cryptographic Hardware and Embedded Systems}.\hskip 1em plus 0.5em minus 0.4em\relax Springer, 2000, pp. 238--251.

\bibitem{xiang2020open}
Y.~Xiang, Z.~Chen, Z.~Chen, Z.~Fang, H.~Hao, J.~Chen, Y.~Liu, Z.~Wu, Q.~Xuan, and X.~Yang, ``Open dnn box by power side-channel attack,'' \emph{IEEE Transactions on Circuits and Systems II: Express Briefs}, vol.~67, pp. 2717--2721, 11 2020.

\bibitem{pedram2006thermal}
M.~Pedram and S.~Nazarian, ``Thermal modeling, analysis, and management in vlsi circuits: Principles and methods,'' \emph{Proceedings of the IEEE}, vol.~94, no.~8, pp. 1487--1501, 2006.

\bibitem{liu2007accurate}
Y.~Liu, R.~P. Dick, L.~Shang, and H.~Yang, ``Accurate temperature-dependent integrated circuit leakage power estimation is easy,'' in \emph{2007 Design, Automation \& Test in Europe Conference \& Exhibition}.\hskip 1em plus 0.5em minus 0.4em\relax IEEE, 2007, pp. 1--6.

\bibitem{vassighi2006thermal}
A.~Vassighi and M.~Sachdev, \emph{Thermal and power management of integrated circuits}.\hskip 1em plus 0.5em minus 0.4em\relax Springer Science \& Business Media, 2006.

\bibitem{arm2015cortexa72}
``Cortex-a72 software optimization guide,'' ARM Ltd., Tech. Rep. ARM UAN 0016A, 2015, available: \url{https://documentation-service.arm.com/static/5ed75eeeca06a95ce53f93c7}.

\bibitem{campagna2019supersingular}
M.~Campagna, C.~Costello, B.~Hess, A.~Jalali, B.~Koziel, B.~LaMacchia, P.~Longa, M.~Naehrig, J.~Renes, D.~Urbanik \emph{et~al.}, ``Supersingular isogeny key encapsulation,'' 2019.

\bibitem{de2022sike}
L.~De~Feo, N.~El~Mrabet, A.~Gen{\^e}t, N.~Kaluderovi{\'c}, N.~L. de~Guertechin, S.~Ponti{\'e}, and {\'E}.~Tasso, ``Sike channels: Zero-value side-channel attacks on sike,'' \emph{IACR Transactions on Cryptographic Hardware and Embedded Systems}, vol. 2022, pp. 264--289, 2022.

\end{thebibliography}
\end{document}